\documentclass[12pt,showpacs,amsmath,amssymb,nofootinbib,superscriptaddress,longbibliography]{book}

\usepackage{graphicx}
\usepackage{epstopdf}
\usepackage{epsfig}
\usepackage{bm}
\usepackage{amsfonts}
\usepackage{amssymb}
\usepackage{color}
\usepackage{float}
\usepackage{mathtools}
\usepackage{amsmath} 
\usepackage{mathpazo}              
\usepackage{dcolumn}
\usepackage{hhline}
\usepackage{hyperref}
\usepackage{array}
\usepackage{lscape}
\usepackage{hyperref}
\usepackage{array}
\usepackage{booktabs}
\usepackage{multirow}
\usepackage[english,greek,french,german]{babel}
\usepackage[utf8x]{inputenc}
\usepackage{txfonts} 
\usepackage{comment}
\usepackage{ mathrsfs }
\usepackage{notoccite}
\usepackage[numbers,sort&compress]{natbib}

\providecommand{\U}[1]{\protect\rule{.1in}{.1in}}

\newcommand{\newc}{\newcommand}

\newc{\be}{\begin{equation}}
\newc{\ee}{\end{equation}}

\newc{\ba}{\begin{eqnarray}}
\newc{\ea}{\end{eqnarray}}
\newc{\bea}{\begin{eqnarray*}}
\newc{\eea}{\end{eqnarray*}}
\newc{\D}{\partial}
\newc{\ie}{{\it i.e.} }
\newc{\eg}{{\it e.g.} }
\newc{\etc}{{\it etc.} }
\newc{\etal}{{\it et al.}}
\newc{\lcdm}{$\Lambda$CDM }
\newcommand{\nn}{\nonumber}
\newc{\ra}{\Rightarrow}
 
\allowdisplaybreaks

\title{\vspace{-8 cm}{\includegraphics[width=4.2cm]{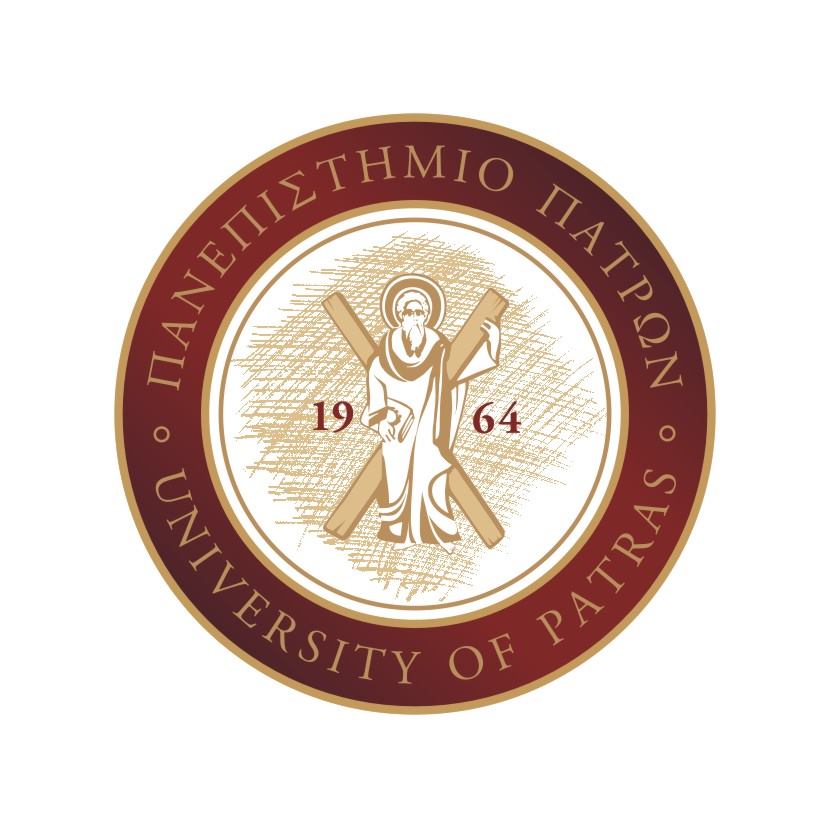}\\ \vspace{0.2cm} \normalsize 
\textlatin{University of Patras}\\
\textlatin{School of Natural Sciences}\\  
\textlatin{Department of Physics}\\
 \textlatin{Division of Theoretical and Mathematical Physics},\\\textlatin{ Astronomy and Astrophysics}\\}  \vspace{2cm} \textbf{\textlatin{An Introduction to FRW Cosmology and dark energy models}}}
\author{{\textlatin{Konstantinos Xenos}} \vspace{1cm} \\ \\ 
{\textlatin{Supervisor: Prof. Smaragda Lola}}
 \\ \vspace{1cm}  }
\date{\textlatin{Thesis submitted to the University of Patras} \\
\textlatin{for the Degree of Undergraduate Studies in Physics} \vspace*{0.5cm} \\
28 09 2020}

\begin{document}

\selectlanguage{english}

\maketitle
\chapter*{Abstract}
In this thesis we will focus on Einstein's interpretation of gravity. We will examine how the most famous equations in cosmology are derived from GR and also some results of cosmological significance. We will see how combining that with observational data forces us to consider some form of dark energy or vacuum energy. So we will conclude with some of the more well-known models for dark energy and examine how the dynamics of dark energy can lead us to the so-called cosmological inflation.

 \tableofcontents

\chapter{Introduction}

In the year 1915 Albert Einstein had finished his 10-year
attempt of constructing the General theory of Relativity(GR).
Almost at the same time the german mathematician David Hilbert had derived the same equations of motion as Einstein via an axiomatic way.
This would come to give space and time the impression of a Rimannian manifold. Thus gravity would be understood as the curvature of spacetime, a now 4-Dimensional hypersurface, due to the existence of mass.

The equations that were derived gave us for the first time in history the capability of studying the universe with the scientific method of mathematics and observation, detaching cosmology from areas such as theoretical astronomy or philosophy.

In the later years this theory has been enriched with knowledge from other areas of physics such as particle physics and quantum field theory, making us wonder whether a Riemannian approach is sufficient for giving us the bigger picture. Moreover, a geometrical approach is not close to illustrating the strong, weak, or even the electromagnetic interactions.

Nevertheless, the fundamental cosmological principles are based upon geometrical terms and Riemannian geometry, even just as a mathematical tool using tools such as the metric or the affine connection , is needed in establishing almost all theories of gravity \cite{Carroll:1997ar,Schutz:1985jx} .

\chapter{The Geometry of curved spacetime}

\section{Einstein's Principle of Equivalence}

The basis of General Relativity is \emph{Einstein's Principle of Equivalence} which equates the inertial and the gravitational mass.
This came to be after a thought experiment of Einstein that connected an accelerating frame of reference with an inertial one under the influence of a gravitational field, where he understood that the physical laws should remain the same for both observers in each frame. This he famously described as $"$the happiest thought of his life$"$ since it paired the gravitational field with the acceleration field and made up the first step for the construction
of a theory for gravity. With Einstein's principle of equivalence we can connect any curved path on an inertial frame to a straight line on an accelerating one, or equally  to a frame under the influence of gravity (those paths are called \emph{geodesics}). We can understand that the spacetime near a gravitational field is not flat since two lines that start with parallel paths don't remain parallel. General Relativity(GR) is a theory of curved spacetime and so we need to work on a geometry of curved surfaces. In contrast to GR the Special theory of Relativity(SR) is a theory of flat spacetime and the space that describes this theory, known as \emph{Minkowski} space, is flat. And this holds true as two world lines that started parallel in the Minkowski space will always remain parallel.

Mathematicians at that time had already been working on  theories of curved space. They were initially proposed by the German mathematician Bernhard Riemann around the year 1856, who was student of Gauss at that time. Riemann was the first one to work on the hypothesis that space does not need to obey Euclid's axioms. That space is known today as \emph{Riemannian} space. It is important to state that Riemannian space can be locally \emph{Euclidian} (flat) but is generally curved (an example of that being a 2-dimensional sphere). The mathematical objects that describe such space are called \emph{Riemannian manifolds}. In short a \emph{topological manifold} of dimension \textbf{n} is a \emph{topological space} that is locally like $\mathbb{R}^{\textbf{n}}$. In addition
a Riemannian manifold is a topological manifold embedded with a \emph{metric}. So in that sense the metric describes any given manifold as we are going to see shortly . 

\section{The metric}

First let us look at the length element $ds^2$ on a curved surface.
For simplicity we are going to work on a 2-dimensional surface existing in $\mathbb{R}^{\textbf{3}}$. Let us consider a position vector $\vec{r}$ to a point $\textbf{P}$ in the surface and then an infinitesimal displacement $d{\vec{r}}$ so that the new position vector of the neighboring  point $\textbf{P}'$ is $\vec{r}'=\vec{r} + d{\vec{r}}$. The length element is going then to be: 

\begin{align*}
ds^2 = d{\vec{r}} \cdot d{\vec{r}}
&=\left(\frac{\partial {\vec{r}}}{\partial x}dx +\frac{\partial {\vec{r}}}{\partial y}dy \right) 
\cdot \left(\frac{\partial {\vec{r}}}{\partial x}dx +\frac{\partial {\vec{r}}}{\partial y}dy \right)  \\
 &= \left(\frac{\partial {\vec{r}}}{\partial x}\right)^2\left(dx\right)^2 +
  2\left(\frac{\partial {\vec{r}}}{\partial x}\right)\left(\frac{\partial {\vec{r}}}{\partial y}\right)dxdy+
 \left(\frac{\partial {\vec{r}}}{\partial y}\right)^2\left(dy\right)^2 \\ 
 &=g_{11}dxdx + 2g_{12}dxdy + g_{22}dydy
\end{align*}\\

\hspace{-0.7cm} The elements $g_{ij}$ then form the matrix:

\begin{equation*}
[g_{ij}]=\begin{pmatrix} g_{11} & g_{12} \\ g_{21} & g_{22} \end{pmatrix}=
\begin{pmatrix}\left(\dfrac{\partial {\vec{r}}}{\partial x}\right)^2 & \dfrac{\partial {\vec{r}}}{\partial x}\dfrac{\partial {\vec{r}}}{\partial y} \\ \dfrac{\partial {\vec{r}}}{\partial x}\dfrac{\partial {\vec{r}}}{\partial y} & \left(\dfrac{\partial {\vec{r}}}{\partial y}\right)^2 \end{pmatrix}
\end{equation*}\\
This is called \emph{the metric tensor} and is a rank 2 tensor. One can also write down the elements of the metric as:

\begin{equation}\label{metric}
g_{ij}=\dfrac{\partial {\vec{r}}}{\partial x^i}\dfrac{\partial {\vec{r}}}{\partial x^j}
\end{equation}\\
Also since the spatial derivatives commute, the metric is a symmetric tensor so:
 
\begin{equation}
g_{ij}=g_{ji}
\end{equation}\\
The upper index indicates the contravariant form of a tensor and the lower index indicates the covariant form. In short a contravariant vector field (rank 1 tensor) transforms by the rule
 
\begin{equation}
\tilde{a}^{\mu}=\frac{\partial \tilde{x}^\mu}{\partial x^\nu}a^{\nu}
\end{equation}\\
and the covariant vector field transforms by the rule

\begin{equation}
\tilde{a}_{\mu}=\frac{\partial x^\nu}{\partial \tilde{x}^\mu}a_{\nu}
\end{equation}

\hspace{-0.7cm} The transformation occurs from the coordinate system $x^{\mu}=(x^1,x^2,...)$ to the \\ $\tilde{x}^{\mu}=(\tilde{x}^1,\tilde{x}^2,...)$ and we are also using the Einstein notation or \emph{Einstein summation convention} where a repeated index would imply a sum (those indices are called dummy indices and they can be changed without  altering the value of the expression). So for a space of dimension \textbf{n} that would be

\begin{equation*}
 a_\mu x^\mu \equiv \sum_{\mu=1}^n a_\mu x^\mu  = a_1x^1 + a_2x^2 + ... + a_nx^n 
\end{equation*}\\
\hspace{-0.7cm}The metric tensor being a rank 2 covariant tensor obeys the transformation rule:

$$\tilde{g}_{ij}=\dfrac{\partial x^\kappa}{\partial \tilde{x}^i}
\dfrac{\partial x^\lambda}{\partial \tilde{x}^j}g_{\kappa\lambda}$$
\\
\hspace{-0.7cm} We can then write down the length element as:

\begin{equation}\label{ds}
ds^2=g_{\mu\nu}dx^{\mu}dx^{\nu}
\end{equation}\\
By electing the metric $g_{\mu\nu}$ to be the Kronecker delta (Euclidean metric) $\delta_{\mu\nu}$ (witch can be easily seen that is also a rank 2 tensor), we obtain that, in a 2-dimensional space where $x^{\mu}=(x,y)$, the length element is:
\begin{equation}
ds^2=\delta_{\mu\nu}dx^{\mu}dx^{\nu}=dx^2+dy^2
\end{equation}

\hspace{-0.7cm}which is Pythagorean theorem. So one can see that the expression \eqref{ds} is a generalization of the Pythagorean theorem on a (generally) curved space. \\ 
\\
\\
\\
\hspace{-0.7cm} Another example can come from SR where the spacetime interval between two events is given by the expression:
\begin{equation*}
ds^2=-(cdt)^2+dx^2+dy^2+dz^2
\end{equation*}
the same expression can be derived by using the metric $\eta_{\mu\nu}=diag(-1,1,1,1)$ on a coordinate system $x^{\mu}=(x^0,x^1,x^2,x^3)=(ct,x,y,z)$. We can now better understand why SR can be described as a theory of 4-dimensional spacetime by using the above notation. That spacetime is called \emph{Minkowski space} and the corresponding metric that defines it \emph{Minkowski metric}. Both the Euclidean and Minkowski metrics describe flat spaces,
the differance being that the Euclidean metric is Riemannian as it has the form\\
$$ g_{\mu\nu}=diag(1,1,...,1)$$
\\
and the Minkowski metric is pseudo-Riemannian as it has the form:

\begin{align*}
 g_{\mu\nu}=diag(-1,1,...,1)
\end{align*}\\
The metric can also be used in order to rise or lower an index, as so 

 \begin{equation*}
g_{\mu\nu} a^{\mu}=a_{\nu}
 \end{equation*} \\
So as its name suggests the General theory of Relativity is indeed a generalization of the Special Theory of relativity as we can examine any given metric $g_{\mu\nu}$
other than $\eta_{\mu\nu}$. In that manner we can further understand Einstein's Principle of Equivalence, saying that every metric $g_{\mu\nu}$ is locally the Minkowski metric $\eta_{\mu\nu}$ and that spacetime is locally Minkowski-like.

\section{The Christoffel symbols}
When working on GR, Einstein realized the importance of working with tensors. As  Einstein's Equivalence Principle states, the laws of physics should be the same for any observer in any coordinate system. Thus, expressing them in terms of tensors is indeed important since tensors are consistent to coordinate transformations. So when we generalize from a flat space to a curved manifold we have to do so in a way which ensures the physical entities having a tensor-like behavior.

In that manner we can see that the ordinary derivative is not behaving as a tensor. Let us consider a transformation

\begin{equation*}
x^{\mu}\rightarrow \tilde{x}^{\mu}=\frac{\partial \tilde{x}^\mu}{\partial x^\nu}x^{\nu}
\end{equation*}\\
For a scalar field $\phi (x^{\mu})$ (rank 0 tensor) we can see that it transforms as

\begin{align*}
\dfrac{\partial \phi (x^{\mu})}{\partial x^\mu} = \partial_{\mu} \phi (x^{\mu}) \rightarrow 
\partial_{\tilde{\mu}} \tilde{\phi} (\tilde{x}^{\mu}) &= \dfrac{\partial}{\partial \tilde{x}^{\mu} } \tilde{\phi} (\tilde{x}^{\mu})  \\
&= \dfrac{\partial x^{\nu}}{\partial \tilde{x}^{\mu}}  \dfrac{\partial}{\partial x^{\nu}} {\phi} (x^{\mu}) \\
&= \dfrac{\partial x^{\nu}}{\partial \tilde{x}^{\mu}}
 \partial_{\nu} \phi (x^{\nu})
\end{align*}\\
so in that case the effect of the ordinary derivative would result in a rank 1 covariant tensor. However this is not the case for higher rank tensors and to prove that we will act the ordinary derivative on a rank 1 tensor, that being the simpler one. For example let us consider the following transformation
\\
\\
\\
\\
\begin{align*}
\partial_{\nu}a_{\mu}(x) \rightarrow \partial_{\tilde{\nu}} \tilde{a_{\mu}}(x) &= \dfrac{\partial x^{\rho}}{\partial \tilde{x}^{\nu}} \partial_{\rho} \tilde{a_{\mu}}(x) \\
&= \dfrac{\partial x^{\rho}}{\partial \tilde{x}^{\nu}} \partial_{\rho} \left(  \dfrac{\partial x^{\kappa}}{\partial \tilde{x}^{\mu}} a_{\kappa}(x)\right) \\
&= \dfrac{\partial x^{\rho}}{\partial \tilde{x}^{\nu}} \partial_{\rho}\dfrac{\partial x^{\kappa}}{\partial \tilde{x}^{\mu}} a_{\kappa}(x) + \dfrac{\partial x^{\rho}}{\partial \tilde{x}^{\nu}} \dfrac{\partial x^{\kappa}}{\partial \tilde{x}^{\mu}} \partial_{\rho} a_{\kappa}(x)
\end{align*} \\
where we can see that because of the first term existing, the quantity $\partial_{\nu}a_{\mu}(x)$ is not a tensor.

So we generalize the concept of the derivative in way that will always result in a tensor. This is called \emph{covariant derivative} and is defined as

\begin{align*}
&\ \nabla_{\nu} \phi = \partial_{\nu} \phi \\
&\ \nabla_{\nu} a_{\mu}= \partial_{\nu} a_{\mu} - \Gamma^{\rho}_{\nu\mu} a_{\rho} \\
&\ \nabla_{\nu} a^{\mu}= \partial_{\nu} a^{\mu} + \Gamma^{\mu}_{\nu\rho} a^{\rho}
\end{align*}\\
The coefficients $\Gamma^{\alpha}_{\beta\gamma}$ are called \emph{connection coefficients} or \emph{Christoffel symbols}.\\
A notation that is usually used for the ordinary and covariant derivative is the following:
 
\begin{equation}\label{notation}
\begin{split}
\partial_\nu a_\mu \equiv a_{\mu , \nu} \\
\nabla_\nu a_{\mu}  \equiv a_{\mu ; \nu}
\end{split}
\end{equation}\\
\\
\\
\\
\\
Given that the Christoffel symbols transform by the law (in a way that does not represent a tensor)

\begin{equation}\label{Christoffeltransformation}
\Gamma^{\alpha}_{\beta\gamma} \rightarrow \dfrac{\partial x^i}{\partial \tilde{x}^\beta}   \dfrac{\partial x^j}{\partial \tilde{x}^\gamma} \dfrac{\partial \tilde{x}^{\alpha}}{\partial x^k} \Gamma^k_{ij} + \dfrac{\partial \tilde{x}^{\alpha}}{\partial x^i} \dfrac{\partial^2 x^i}{\partial \tilde{x}^{\alpha}\partial \tilde{x}^{\beta}}
\end{equation} \\
we can get that the term $\nabla_\nu a_{\mu}$ transforms as

\begin{equation*}
\nabla_\nu a_{\mu} \rightarrow \dfrac{\partial x^i}{\partial \tilde{x}^\nu}   \dfrac{\partial x^j}{\partial \tilde{x}^\mu} \nabla_i a_j
\end{equation*}\\
which defines a rank 2 covariant tensor. This can be generalized to tensors of higher rank and the application  of the covariant derivative to any tensor will increase its covariant rank by one.

The Christoffel symbols can be defined as

\begin{equation}
\dfrac{\partial {e_{\alpha}}}{\partial x^{\beta}}=\partial_{\beta}e_{\alpha}=\Gamma^{\gamma}_{\beta\alpha}e_{\gamma}=\Gamma^{\gamma}_{\alpha\beta}e_{\gamma}
\end{equation}\\
where we assume a torsion-less manifold so that the Christoffel symbols are symetric in regard to the lower indices. We can then write the following form for the metric according to the basis vectors:

\begin{equation}
g_{\alpha\beta}=e_{\alpha} \cdot e_{\beta} 
\end{equation}\\
which derives out of \eqref{metric}. This is also a generalization of the usual connection of the basis vectors, the most usual being the one of a Euclidean space 

\begin{equation*}
e_{\alpha} \cdot e_{\beta}={\delta}_{\alpha\beta} 
\end{equation*}\\
and the one that can be find in SR

\begin{equation*}
e_{\alpha} \cdot e_{\beta}={\eta}_{\alpha\beta} 
\end{equation*}\\
We can now write down the partial derivative of the metric

\begin{align*}
\partial_{\gamma}g_{\alpha\beta} &= \partial_{\gamma}\left(e_{\alpha} \cdot e_{\beta}\right)  \\
&=\left(\partial_{\gamma}e_{\alpha}\right)e_{\beta} +
 e_{\alpha}\left(\partial_{\gamma}e_{\beta}\right) \\
 &= \Gamma^{\delta}_{\alpha\gamma}e_{\delta}e_{\beta} +
e_{\alpha} \Gamma^{\delta}_{\beta\gamma}e_{\delta} \\
&= g_{\delta\beta}\Gamma^{\delta}_{\alpha\gamma} +
 g_{\alpha\delta}\Gamma^{\delta}_{\beta\gamma}
\end{align*}\\
And by combining similar expressions in the following matter we get

\begin{equation}
\partial_{\gamma}g_{\alpha\beta} + \partial_{\beta}g_{\gamma\alpha}- \partial_{\alpha}g_{\beta\gamma} = 
2g_{\alpha\delta}\Gamma^{\delta}_{\beta\gamma}
\end{equation}\\
From that we can now get the following useful formula for the Christoffel \\ symbols
 
\begin{equation}\label{Chrisoffel}
\Gamma^{\delta}_{\beta\gamma} = \frac{1}{2}g^{\alpha\delta}\left(\partial_{\gamma}g_{\alpha\beta} + \partial_{\beta}g_{\gamma\alpha}- \partial_{\alpha}g_{\beta\gamma}\right)
\end{equation}\\
where we use the inverse metric defined as

\begin{equation}\label{invmetric}
g^{\mu\kappa}g_{\kappa\nu}={\delta}^{\mu}_{\nu}
\end{equation}\\
and given that the metric is symmetric we can conclude that the inverse metric is also symmetric.

The same formula for the Christoffel symbols can be also extracted if one notices a very significant property of the metric 
\begin{equation}\label{compatibility}
\nabla_{\rho} g_{\mu\nu} = 0
\end{equation}
This is also called \emph{metric compatibility property} for the covariant derivative operator, and in GR it plays an important role on the development of some models, the most famous being that of the \emph{cosmological constant}.

\section{The Riemman curvature tensor and the Bianchi identities}

We are now going to define an entity that gives a mathematical description of the curvature of a manifold. This is the \emph{Riemann curvature tensor} and is defined as such

\begin{equation}\label{commute}
[\nabla_\mu ,\nabla_\nu]a_\rho = R^{\kappa}_{\rho\nu\mu}a_\kappa 
\end{equation}\\
In short, we can understand the commutator of the covariant derivatives as the measure of the parallel transport of a vector field (here $a$) alongside two different pathways. So in the case of $R^{\kappa}_{\rho\nu\mu}=0$ we would have a flat manifold.

The analytic form of the Riemann tensor comes from expanding the above commutator

\begin{align*}
[\nabla_\mu ,\nabla_\nu]a_\rho &= 
\nabla_\mu (\nabla_\nu a_\rho ) - \nabla_\nu \left(\nabla_\mu a_\rho \right) 
\end{align*}\\
where we can treat the term $\nabla_\nu a_\rho$ as a rank 2 covariant tensor and write its \\ covariant derivative as 

\begin{align*}
\nabla_\mu (\nabla_\nu a_\rho ) = \partial_\mu (\nabla_\nu a_\rho ) -
\Gamma^{\lambda}_{\mu\nu}(\nabla_\lambda a_\rho)  -
\Gamma^{\lambda}_{\mu\rho}(\nabla_\nu a_\lambda)
\end{align*}\\
so the above expression becomes

\begin{align*}
[\nabla_\mu ,\nabla_\nu]a_\rho &=
 \partial_\mu (\nabla_\nu a_\rho ) -
\Gamma^{\lambda}_{\mu\nu}(\nabla_\lambda a_\rho)  -
\Gamma^{\lambda}_{\mu\rho}(\nabla_\nu a_\lambda) \\
&\ - \partial_\nu (\nabla_\mu a_\rho ) +
\Gamma^{\lambda}_{\nu\mu}(\nabla_\lambda a_\rho)  +
\Gamma^{\lambda}_{\nu\rho}(\nabla_\mu a_\lambda)\\
&=  \partial_\mu (\partial_\nu a_\rho - \Gamma^{\kappa}_{\nu\rho}a_\kappa) - 
\Gamma^{\lambda}_{\mu\rho}(\partial_\nu a_\lambda -
 \Gamma^{\kappa}_{\nu\lambda}a_\kappa) \\
&\ - \partial_\nu (\partial_\mu a_\rho - \Gamma^{\kappa}_{\mu\rho}a_\kappa) + 
\Gamma^{\lambda}_{\nu\rho}(\partial_\mu a_\lambda - \Gamma^{\kappa}_{\mu\lambda}a_\kappa) \\
&= -\partial_\mu (\Gamma^{\kappa}_{\nu\rho})a_\kappa -
 \Gamma^{\kappa}_{\nu\rho}\partial_\mu a_\kappa -
 \Gamma^{\lambda}_{\mu\rho}\partial_\nu a_\lambda +
 \Gamma^{\lambda}_{\mu\rho}\Gamma^{\kappa}_{\nu\lambda}a_\kappa \\ 
 &\ + \partial_\nu (\Gamma^{\kappa}_{\mu\rho})a_\kappa +
 \Gamma^{\kappa}_{\mu\rho}\partial_\nu a_\kappa +
 \Gamma^{\lambda}_{\nu\rho}\partial_\mu a_\lambda -
 \Gamma^{\lambda}_{\nu\rho}\Gamma^{\kappa}_{\mu\lambda}a_\kappa
\end{align*}
We can now observe that some terms cancel each other out

\begin{equation*}
\Gamma^{\lambda}_{\nu\rho}\partial_\nu a_\lambda - 
\Gamma^{\kappa}_{\nu\rho}\partial_\mu a_\kappa = 0 
\quad\text{and}\quad 
\Gamma^{\lambda}_{\mu\rho}\partial_\nu a_\lambda - 
\Gamma^{\kappa}_{\mu\rho}\partial_\mu a_\kappa = 0 
\end{equation*}\\
since $\kappa ,\lambda$ are dummy indices. So we end up with

\begin{align*}
[\nabla_\mu ,\nabla_\nu]a_\rho &=
\left[\partial_\nu (\Gamma^{\kappa}_{\mu\rho}) - \partial_\mu (\Gamma^{\kappa}_{\nu\rho}) +
\Gamma^{\lambda}_{\mu\rho}\Gamma^{\kappa}_{\nu\lambda} - \Gamma^{\lambda}_{\nu\rho}\Gamma^{\kappa}_{\mu\lambda}\right]a_\kappa
\end{align*}\\
and comparing with the definition of the Riemann tensor \eqref{commute} we get the\\ formula
\begin{equation}\label{Riemann}
R^{\kappa}_{\rho\nu\mu} = \partial_\nu (\Gamma^{\kappa}_{\mu\rho}) - \partial_\mu (\Gamma^{\kappa}_{\nu\rho}) +
\Gamma^{\lambda}_{\mu\rho}\Gamma^{\kappa}_{\nu\lambda} - \Gamma^{\lambda}_{\nu\rho}\Gamma^{\kappa}_{\mu\lambda}
\end{equation}
or by using the notation \eqref{notation}

\begin{equation}\label{Riemann2}
R^{\kappa}_{\rho\nu\mu} = \Gamma^{\kappa}_{\mu\rho ,\nu} -  \Gamma^{\kappa}_{\nu\rho ,\mu} +
\Gamma^{\lambda}_{\mu\rho}\Gamma^{\kappa}_{\nu\lambda} - \Gamma^{\lambda}_{\nu\rho}\Gamma^{\kappa}_{\mu\lambda}
\end{equation}\\
The Riemann tensor is a rank 4 mixed tensor of type (1,3) and we can make it a rank 4 covariant tensor by applying the metric $g_{\mu\nu}$: 

\begin{equation}\label{Riemann3}
R_{\sigma\rho\nu\mu} \equiv
 g_{\sigma\kappa}R^{\kappa}_{\rho\nu\mu}
\end{equation}\\
and we can get some useful properties of the Riemann tensor using the above formulas

\begin{align}
&\ R^{\kappa}_{\rho\nu\mu} = -R^{\kappa}_{\rho\mu\nu}
\quad\text{or}\quad 
R^{\kappa}_{\rho [\nu\mu]} = 0 \label{Riemann4} \\ \nn \\
&\ R^{\kappa}_{\rho\nu\mu}  +
 R^{\kappa}_{\nu\mu\rho} +
  R^{\kappa}_{\mu\rho\nu} = 0 
  \quad\text{or}\quad
  R_{\sigma\rho\nu\mu}  +
 R_{\sigma\nu\mu\rho} +
  R_{\sigma\mu\rho\nu} = 0     \label{Riemann5} \\ \nn \\
&\ R_{\sigma\rho\nu\mu} = 
R_{\nu\mu\sigma\rho} =  
-R_{\rho\sigma\nu\mu} =
-R_{\sigma\rho\mu\nu} =
R_{\rho\sigma\nu\mu}\label{Riemann6}
\end{align}\\
and the more significant identity

\begin{align}
\nabla_\lambda R_{\sigma\rho\nu\mu } +
   \nabla_\nu R_{\sigma\rho\mu\lambda } +
   \nabla_\mu R_{\sigma\rho\lambda \nu} \equiv
 R_{\sigma\rho\nu\mu ;\lambda} +
     R_{\sigma\rho\mu\lambda ;\nu} + 
     R_{\sigma\rho\lambda \nu;\mu} = 0 \label{Bianchi1}
\end{align}\\
or in a more compact way

\begin{equation}
R_{\sigma\rho[\nu\mu ;\lambda]} =0 \label{Bianchi2}
\end{equation}\\
These identities are also known as \emph{the Bianchi identities for the Riemann tensor} and they too play a significant role in establishing some important equations in GR and in cosmology.

\section{The Ricci tensor}

The \emph{Ricci tensor} is defined as the contraction of the Riemann curvature tensor

\begin{equation}\label{Ricci}
R_{\mu\nu} \equiv R^\kappa_{\;\mu\kappa\nu} =
\partial_\kappa (\Gamma^{\kappa}_{\;\mu\nu}) - \partial_\nu (\Gamma^{\kappa}_{\;\kappa\mu}) +
\Gamma^{\lambda}_{\;\nu\mu}\Gamma^{\kappa}_{\;\kappa\lambda} - \Gamma^{\lambda}_{\;\kappa\mu}\Gamma^{\kappa}_{\;\nu\lambda}
\end{equation}\\
We can consider other contractions for the Ricci tensor such as $R^\kappa_{\;\kappa\mu\nu}$, but using the anti-symmetry properties of the Riemman tensor we end up with

\begin{align*}
 R_{\sigma\rho\nu\mu} = - R_{\rho\sigma\nu\mu} &\Rightarrow 
  g^{\sigma\rho}R_{\sigma\rho\nu\mu}  = - g^{\sigma\rho} R_{\rho\sigma\nu\mu} \\
&\ \Rightarrow  R^{\rho}_{\rho\nu\mu} = -R^{\rho}_{\rho\nu\mu}  \\
&\ \Rightarrow  R^{\rho}_{\rho\nu\mu} = 0
\end{align*}\\
In a similar way we find that the contraction $R^\lambda_{\mu\nu\lambda}=-R^\lambda_{\mu\lambda\nu}$ (also trivial) and by using a symmetry property for the Riemann tensor we can show that the Ricci tensor is symmetric

\begin{equation}
R_{\mu\nu}=R_{\nu\mu}
\end{equation}\\
This symmetry implies that for a 4-dimensional space the independent components for the Ricci tensor are 10.
\\
\\
\\
\\
Furthermore by applying the metric to the Ricci tensor we define the \emph{Ricci scalar}

\begin{equation}\label{Ricciscalar}
R \equiv R^{\mu}_{\;\mu} = g_{\mu\nu}R^{\mu\nu}
\end{equation}\\
and as we can see it is invariant under coordinate transformations, a property that makes it very useful in perceiving the magnitude of curvature for a given manifold.

\section{The Einstein tensor}

We can now use the Bianchi identities \eqref{Bianchi1} to define the \emph{Einstein tensor}. We apply the metric tensor as following

\begin{align*}
&\   g^{\sigma\nu}g^{\rho\mu}\nabla_\lambda R_{\sigma\rho\nu\mu} +
 g^{\sigma\nu}g^{\rho\mu}  \nabla_\nu R_{\sigma\rho\mu\lambda } +
  g^{\sigma\nu}g^{\rho\mu} \nabla_\mu R_{\sigma\rho\lambda \nu} = 0
\end{align*}\\
and since $\nabla_\kappa g_{\mu\nu} = 0$ we can insert the metric tensors to the covariant derivatives

\begin{align*}
 \nabla_\lambda(g^{\sigma\nu}g^{\rho\mu} R_{\sigma\rho\nu\mu}) +
\nabla_\nu ( g^{\sigma\nu}g^{\rho\mu}  R_{\sigma\rho\mu\lambda}) +
  \nabla_\mu(g^{\sigma\nu}g^{\rho\mu}  R_{\sigma\rho\lambda \nu}) = 0
\end{align*}\\
now we notice that for the first term

\begin{align*}
g^{\sigma\nu}R_{\sigma\rho\nu\mu} \equiv 
R^{\nu}_{\rho\nu\mu} \equiv 
R_{\rho\mu} 
\quad\text{and}\quad 
g^{\rho\mu} R_{\rho\mu} \equiv R
\end{align*}\\
and for the other two terms we use the anti-symmetry properties for the Riemann tensor \eqref{Riemann6} as so

\begin{align*}
R_{\sigma\rho\mu\lambda} = - R_{\rho\sigma\mu\lambda} 
\quad\text{and}\quad
R_{\sigma\rho\lambda \nu} = - R_{\sigma\rho\nu\lambda}
\end{align*}\\
so we get

\begin{align*}
\nabla_\lambda R -
 \nabla_\nu(g^{\sigma\nu}g^{\rho\mu} R_{\rho\sigma\mu\lambda}) -
 \nabla_\mu(g^{\sigma\nu}g^{\rho\mu} R_{\sigma\rho\nu\lambda})=0
\end{align*}\\
where similarly we can contract the last two terms

\begin{align*}
g^{\rho\mu} R_{\rho\sigma\mu\lambda} = R_{\sigma\lambda}
\quad\text{and}\quad
g^{\sigma\nu}R_{\sigma\rho\nu\lambda} = R_{\rho\lambda}
\end{align*}\\
and then the  previous expression is just

\begin{align*}
\nabla_\lambda R - 
\nabla_\nu(g^{\sigma\nu} R_{\sigma\lambda}) -
\nabla_\mu(g^{\rho\mu}R_{\rho\lambda}) = 0
\end{align*}\\
but by  noticing that the last two terms are the same if we swap the dummy indices $(\sigma \leftrightarrow \rho)$ ,
 $(\mu \leftrightarrow \nu)$ we get
 
\begin{align*}
 \nabla_\lambda R - 
  2 \nabla_\nu(g^{\sigma\nu} R_{\sigma\lambda})=0 
\end{align*}\\
which is the same as

\begin{align*}
\nabla_\nu(\delta^{\nu}_{\;\lambda} R - 2g^{\sigma\nu} R_{\sigma\lambda})=0
 \intertext{\centering or} 
 \nabla_\nu(g^{\sigma\nu} R_{\sigma\lambda} -
  \dfrac{1}{2} \delta^{\nu}_{\;\lambda} R) =0
\end{align*}\\
So we end up with

\begin{equation}\label{Bianchi3}
\nabla_\nu(R^{\nu}_{\;\lambda} - \dfrac{1}{2} \delta^{\nu}_{\;\lambda} R) \equiv
(R^{\nu}_{\lambda} - \dfrac{1}{2} \delta^{\nu}_{\;\lambda} R)_{;\nu}=0
\end{equation}\label{EinsteinTensor}\\
And now we can define the \emph{Einstein tensor} as

\begin{equation}
G^{\mu}_{\;\nu} \equiv R^{\mu}_{\;\nu} - \dfrac{1}{2} \delta^{\mu}_{\;\nu} R 
\quad\text{or}\quad
 G_{\mu\nu} \equiv R_{\mu\nu} - \dfrac{1}{2}g_{\mu\nu} R
\end{equation}\\
And form \eqref{Bianchi3} see that it satisfies the Bianchi identities as so

\begin{equation}\label{BianchiEinstein}
\nabla_{\mu} G^{\mu}_{\;\nu} \equiv G^{\mu}_{\;\nu ;\mu} =0 \\
\quad\text{or}\quad
\nabla_{\mu} G^{\mu\nu} \equiv G^{\mu\nu}_{\;\;\;;\mu} =0
\end{equation}\\
And since both the Ricci tensor and the metric that compose the Einstein tensor are symmetric, the Einstein tensor in turn
is symmetric as well

\begin{equation}
G_{\mu\nu} = G_{\nu\mu}
\end{equation}\\
\\
The Einstein tensor plays an important role in the connection of geometry and gravity as we will see later on, and the fact that it satisfies the Bianchi identities \eqref{BianchiEinstein} is also of significance.

\chapter{General Relativity}

\section{The action principle}

The \emph{principle of the least action} or \emph{Hamilton's principle} states that the evolution of a system will occur in a way that the action between two states is stationary for small variations of the variables. In mathematical terms this means that for the action

\begin{equation}
S \equiv \int_{t_1}^{t_2} \mathscr{L}(q(t),\dot q(t),t)dt
\end{equation}\\
Hamilton's principle demands the following condition:

\begin{equation}
\dfrac{\delta S}{\delta q} =0
\end{equation}\\
where $\mathscr{L}(q(t),\dot q(t),t)$ is the Lagrangian function and the q(t) are the generalized coordinates.
\\
\\
\\
We can go a step further and write the Lagrangian as so

\begin{equation}
\mathscr{L}(q(t),\dot q(t),t) \equiv \int \mathcal{L}(q(t),\dot q(t),t)d^nx
\end{equation}\\
where the $\mathcal{L}$ is called \emph{Lagrangian density}. Then the action can be written as

\begin{equation}
S = \int \mathcal{L}(q(t),\dot q(t),t)d^{n}x dt
\end{equation}\\

So by using the appropriate action we can derive various equations of motion via this variation approach. This is called the \emph{Lagrangian formalism}

\section{The Hilbert-Einstein action}

The action for spacetime curvature that was proposed by Hilbert in order to derive Einstein's field equations is known as the Hilbert-Einstein action

\begin{equation}\label{HilbertAction}
S_{\scriptscriptstyle H-E}= \frac{c^4}{16{\pi}G_{\scriptscriptstyle N}}\int{R\sqrt{-g}d^{4}x}
\end{equation}\\
or by noting a constant $k=\cfrac{8{\pi}G_{\scriptscriptstyle N}}{c^4}$ , also known as Einstein's constant

\begin{equation}
S_{\scriptscriptstyle H-E}= \frac{1}{2k}\int{R\sqrt{-g}d^{4}x}
\end{equation}\\
(although we are not going to follow this notation, since the same symbol is used for the curvature of spacetime which tends to appear more often in cosmology). \\

The term $\sqrt{-g}$ is added so that the volume element $d^{4}x=dx^0dx^1dx^2dx^3$ is generalized to the volume element of a 4-dimensional topological manifold, where $g=det(g_{\mu\nu})$ and can be seen as the Jacobian of a transformation from the ordinary coordinate system to that of a curved spacetime. Also the $"\text{--}"$ derives from the fact that in General Relativity we usually are occupied with pseudo-Riemannian metrics. The most simple example of a pseudo-Riemannian metric is the Minkowski metric ${\eta}_{\mu\nu}=diag(-1,1,1,1)$ where in this example we can see that $det({\eta}_{\mu\nu})= -1$ so the $"\text{--}"$ ensures that the quantity under the root is positive.

The reasoning behind the election of the Lagrangian density is pretty straightforward. We want the Lagrangian density to be a scalar that consists of the metric and its derivatives, the simplest one being the Ricci scalar R. One can suggest other scalars that describe the curvature to be the Lagrangian density, like $R^{\mu\nu}R_{\mu\nu}$ or even $R^{\kappa\lambda\mu\nu}R_{\kappa\lambda\mu\nu}$ , in correspondence to the electromagnetic Lagrangian 
$F^{\mu\nu} F_{\mu\nu}$. In fact there is indeed work being done towards that direction, constructing theories of generalized gravity, but we are not going to concern with any of those in this thesis.

\section{The Einstein field equations}

We are now going to use the variation approach in order to derive the Einstein field equations by using the Hilbert-Einstein action \eqref{HilbertAction}. Since the Ricci scalar is a function of the metric it's natural that the variation of the action is going to be in regard of the the metric. So the condition is the following:

\begin{equation}\label{condition1}
\dfrac{\delta S_{\scriptscriptstyle H-E}}{\delta g^{\mu\nu}}=0
\end{equation}\\
\\
\\
\\
So by developing the term $\delta S_{\scriptscriptstyle H-E}$ we get

\begin{align}
\delta S_{\scriptscriptstyle H-E} = 
\delta \left(\frac{c^4}{16{\pi}G_{\scriptscriptstyle N}}\int{R\sqrt{-g}d^{4}x}\right)
\end{align}\\
and for now we can ignore the constant and just work with the integral

\begin{align}
\delta \left(\int{R\sqrt{-g}d^{4}x}\right) &=
\int \delta(R\sqrt{-g})d^{4}x
\end{align}\\
where we used a property of the variation (not in a strict mathematical manner)

\begin{align}
\delta \left(\int f\right) = \int \delta f
\end{align}\\
and, since its not so uncommon, we are going to treat the variation as an ordinary differential, by applying a similar product rule
\begin{align}
\delta (fg) = g\delta f + f\delta g
\end{align}\\
So by expanding the Ricci scalar, while using the chain rule, we get 

\begin{align*}
\delta(R\sqrt{-g}) &=
\delta(g^{\mu\nu}R_{\mu\nu}\sqrt{-g}) \\
&=  R_{\mu\nu}\sqrt{-g} \delta g^{\mu\nu} +
   g^{\mu\nu}R_{\mu\nu} \delta \sqrt{-g}+
   \sqrt{-g}g^{\mu\nu}\delta R_{\mu\nu} \\
&=  R_{\mu\nu}\sqrt{-g} \delta g^{\mu\nu}+
   R\delta \sqrt{-g} +
   \sqrt{-g}g^{\mu\nu}\delta R_{\mu\nu}
\end{align*}\\
Now we only have to calculate the expressions 
$\delta \sqrt{-g}$ and $\delta R_{\mu\nu}$. For the first one, we use the following result from linear algebra, regarding a given square matrix 

\begin{align}
adj(A)=det(A)\cdot inv(A) \quad\text{or}\quad 
A^{-1} = \dfrac{1}{det(A)}C^\top
\end{align}\\
where the \emph{adj(A)} is the adjugate of A, meaning 

\begin{align}
adj(A) = C^\top
\end{align}\\
if C is the cofactor matrix of A, but we are not going to go into more detail. Applying this for the metric tensor we get

\begin{align}
inv[g_{\mu\nu}] = \dfrac{adj[g_{\mu\nu}]}{g} =
\dfrac{[G^{\mu\nu}]^\top}{g}
\end{align}\\
$G^{\mu\nu}$ is not the Einstein tensor but the cofactor of the metric tensor and after multiplying the above expression by
 $[g_{\mu\nu}]$ we get that 
 
\begin{align}\label{cofactor}
[g_{\mu\nu}]\dfrac{[G^{\mu\nu}]^\top}{g}=I 
\end{align}\\
and since the metric tensor is symmetric, its cofactor would also be symmetric, and thus equation \eqref{cofactor} becomes

\begin{align}\label{cofactor2}
 g_{\mu\nu}\dfrac{G^{\mu\nu}}{g} = \delta^{\nu}_{\mu}
 \quad\text{or}\quad
 g_{\mu\nu}G^{\mu\nu}=g
\end{align}\\
and we can easily connect the above by the inverse metric if we define it as

\begin{align}
g^{\mu\nu} \equiv \dfrac{G^{\mu\nu}}{g} 
\end{align}\\
which agrees with the previous definition of then inverse metric \eqref{invmetric} \\
We can then use \eqref{cofactor2} to write down the cofactor of the metric in a useful way that allows us to get an expression for the $\delta g$

\begin{align}
G^{\mu\nu}=\dfrac{\partial g}{\partial g_{\mu\nu}}
\end{align}\\
and so the variation of g would be

\begin{equation}\label{deltag1}
\delta g = G^{\mu\nu}\delta g_{\mu\nu} = 
g g^{\mu\nu} \delta g_{\mu\nu}
\end{equation}\\
and in a similar way in regard to $\delta g^{\mu\nu}$ it would be

\begin{equation}\label{deltag2}
\delta g = - g g_{\mu\nu} \delta g^{\mu\nu}
\end{equation}\\
From this we can use the chain rule to calculate $\delta \sqrt{-g}$

\begin{equation}\label{deltag3}
\begin{split}
\delta \sqrt{-g} =- \dfrac{\sqrt{-g}}{g} \delta g
&= - \dfrac{1}{2\sqrt{-g}}(-g g_{\mu\nu} \delta g^{\mu\nu})\\
&= - \dfrac{1}{2}\sqrt{-g}g_{\mu\nu} \delta g^{\mu\nu}
\end{split}
\end{equation}\\
\\
\\
And so the initial variation becomes

\begin{equation}
\begin{split}
\delta(R\sqrt{-g}) &=
   R_{\mu\nu}\sqrt{-g} \delta g^{\mu\nu}+
   R \text{ } \delta  \sqrt{-g} +
   \sqrt{-g}g^{\mu\nu}\delta R_{\mu\nu} \\  
   &= R_{\mu\nu}\sqrt{-g} \delta g^{\mu\nu} -
  \dfrac{\sqrt{-g}}{2} g_{\mu\nu} R \delta g^{\mu\nu} +
   \sqrt{-g}g^{\mu\nu}\delta R_{\mu\nu} \\
   &= \sqrt{-g}\left(R_{\mu\nu} - \dfrac{1}{2} g_{\mu\nu} R \right) \delta g^{\mu\nu} +
   \sqrt{-g}g^{\mu\nu}\delta R_{\mu\nu}
\end{split}
\end{equation}\\
The next thing should be to express the $\delta R_{\mu\nu}$ term in regard to $\delta g^{\mu\nu}$, but we are going to show that this term's contribution to the action is actually zero. The easiest way to accomplish 
that is by using an identity that connects the $\delta R_{\mu\nu}$ with the Christoffel symbols. This is known as the \emph{Palatini identity}

\begin{equation}\label{Palatini}
\delta R_{\mu\nu} = \nabla_\rho(\delta \Gamma^\rho_{\mu\nu}) - 
\nabla_\nu(\delta \Gamma^\rho_{\rho\mu})
\end{equation}\\
The proof of it is pretty straightforward

\begin{align*}
\nabla_\rho(\delta \Gamma^\rho_{\mu\nu}) - 
\nabla_\nu(\delta \Gamma^\rho_{\rho\mu}) &= 
\partial_\rho \delta \Gamma^\rho_{\mu\nu} +
 \Gamma^\rho_{\rho\kappa}\delta \Gamma^\kappa_{\mu\nu} -
 \Gamma^\kappa_{\rho\nu}\delta \Gamma^\rho_{\mu\kappa} -
 \Gamma^\kappa_{\rho\mu}\delta \Gamma^\rho_{\kappa\nu} \\
 &\quad-\left(\partial_\nu \delta \Gamma^\rho_{\rho\mu} +
 \Gamma^\rho_{\nu\kappa}\delta \Gamma^\kappa_{\rho\mu} -
 \Gamma^\kappa_{\nu\rho}\delta \Gamma^\rho_{\kappa\mu}-
 \Gamma^\kappa_{\nu\mu}\delta \Gamma^\rho_{\rho\kappa}\right)\\
 &=\partial_\rho \delta \Gamma^\rho_{\mu\nu} -
 \partial_\nu \delta \Gamma^\rho_{\rho\mu} +
 \delta(\Gamma^\kappa_{\nu\mu} \Gamma^\rho_{\rho\kappa})
 -\delta(\Gamma^\rho_{\nu\kappa} \Gamma^\kappa_{\rho\mu})\\
 &=\delta \left(\partial_\rho \Gamma^\rho_{\mu\nu} -
 \partial_\nu \Gamma^\rho_{\rho\mu} +
 \Gamma^\kappa_{\nu\mu} \Gamma^\rho_{\rho\kappa}
 -\Gamma^\rho_{\nu\kappa} \Gamma^\kappa_{\rho\mu}\right)\\
 &=\delta R_{\mu\nu}
\end{align*}\\
\\
\\
\\
An issue that can come up is that since the Christoffel symbols are no tensors, then one should not be able to write down the covariant derivative in such manner. But since we take a small variation, one can ignore the non linear term that appears on \eqref{Christoffeltransformation} and thus the variation of the Christoffel symbols can be treated as a tensor.\\

\hspace{-0.65cm} Using this result we can now calculate even further

\begin{align*}
g^{\mu\nu}\delta R_{\mu\nu} &=
 g^{\mu\nu} \left(\nabla_\rho(\delta \Gamma^\rho_{\mu\nu}) - 
\nabla_\nu(\delta \Gamma^\rho_{\rho\mu})\right) \\
&= g^{\mu\nu}\nabla_\rho(\delta \Gamma^\rho_{\mu\nu}) -
g^{\mu\rho}\nabla_\rho(\delta \Gamma^\rho_{\rho\mu}) \\
&=\nabla_\rho \left( g^{\mu\nu}\delta \Gamma^\rho_{\mu\nu} -
 g^{\mu\rho}\delta \Gamma^\rho_{\rho\mu}\right) - 
 \left(\delta \Gamma^\rho_{\mu\nu}\nabla_\rho(g^{\mu\nu}) -
 \delta \Gamma^\rho_{\rho\mu} \nabla_\rho ( g^{\mu\rho})\right) \\
 &=\nabla_\rho \left( g^{\mu\nu}\delta \Gamma^\rho_{\mu\nu} -
 g^{\mu\rho}\delta \Gamma^\rho_{\rho\mu}\right)
\end{align*}\\
since the second term is zero due to the metric compatibility \eqref{compatibility}. So the contribution to the action is the integral

\begin{equation}
\int \sqrt{-g} d^4 x g^{\mu\nu}\delta R_{\mu\nu} =
 \int \sqrt{-g} d^4 x \nabla_\rho 
 \left( g^{\mu\nu}\delta \Gamma^\rho_{\mu\nu} -
 g^{\mu\rho}\delta \Gamma^\rho_{\rho\mu}\right)
\end{equation}\\
which is an integral over a total derivative, meaning it results to the boundary terms. We assume however that for a realistic field those terms are going to vanish, as we go to infinity. Thus this term is going to be zero. So we end up with

\begin{equation}
\delta S_{\scriptscriptstyle H-E}= 
\int\sqrt{-g} d^4 x\left(R_{\mu\nu} -
 \dfrac{1}{2} g_{\mu\nu} R \right) \delta g^{\mu\nu} 
\end{equation}\\
\\
\\
Then the condition \eqref{condition1} gives us the field equations in vacuum or in the absence of mass and other sources

\begin{equation}\label{field0}
R_{\mu\nu} -
 \dfrac{1}{2} g_{\mu\nu} R =0 
 \quad\text{or}\quad G_{\mu\nu}=0
\end{equation}\\
The above equation holds true if and only if $R_{\mu\nu}=0$, meaning a Ricci-flat space. An obvious solution to this is the case of the Minkowski metric $\eta_{\mu\nu}$ since in that case
$
\Gamma^{\alpha}_{\;\mu\nu}=0 \;\;\forall\; \mu,\nu,\alpha 
$
as the partial derivatives of the metric all vanish. This solution leads us however to a trivial case of $R^\kappa_{\mu\nu\sigma} = 0$ which means a Riemann-flat space. The non-trivial solutions are those in which the Riemann tensor is non-zero but the Ricci tensor is. \\

Next, in order to derive the non-vacuum field equations, we have to assume the addition of another action which will depend in the form of the source. In the case of matter we assume an action for matter

\begin{equation}
S_{\scriptscriptstyle M} =
 \int \mathcal{L_{\scriptscriptstyle M}} \sqrt{-g} d^4x
\end{equation}\\
where $\mathcal{L_{\scriptscriptstyle M}}$ is the corresponding Lagrangian for the contribution of matter.\\
Then the total action is going to be

\begin{equation}
S_{tot}= 
S_{\scriptscriptstyle H-E} +
S_{\scriptscriptstyle M}
\end{equation}\\
and the condition from the action principle is going to be 

\begin{equation}
\dfrac{\delta S_{tot} }{\delta g^{\mu\nu}}=0
\end{equation}\\
The expression for $\delta S_{tot}$ after ignoring the term $\delta R_{\mu\nu}$
is going to be

\begin{align*}
\delta S_{tot} &=
 \delta \left(\frac{c^4}{16{\pi}G}\int R\sqrt{-g}\;d^4x +
  \int \mathcal{L_{\scriptscriptstyle M}} \sqrt{-g}\; d^4x\right)\\
  &=\int \left[\frac{c^4}{16{\pi}G}\sqrt{-g}\left(R_{\mu\nu} -
   \dfrac{1}{2}g_{\mu\nu} R \right)\delta g^{\mu\nu}+
  \sqrt{-g}\; \delta \mathcal{L_{\scriptscriptstyle M}} + 
  \mathcal{L_{\scriptscriptstyle M}}\; \delta\sqrt{-g}
   \right] d^4x
\end{align*}\\
And so the condition reads as follows

\begin{equation}\label{condition3}
\frac{c^4}{16{\pi}G}\left(R_{\mu\nu} - \dfrac{1}{2}g_{\mu\nu} R \right) + 
   \frac{1}{\sqrt{-g}}\dfrac{\delta S_{\scriptscriptstyle M}}{\delta g^{\mu\nu}}=0
\end{equation}\\
\begin{flalign*}
\text{where}\quad \delta S_{\scriptscriptstyle M} = \int d^4x \left(  
\sqrt{-g}\; \delta \mathcal{L_{\scriptscriptstyle M}} - 
 \frac{1}{2} \sqrt{-g} g_{\mu\nu} \mathcal{L_{\scriptscriptstyle M}} \delta g^{\mu\nu} \right) &&
\end{flalign*}\\
After we define the energy-momentum tensor as

\begin{equation}
T_{\mu\nu} \equiv - \dfrac{2}{\sqrt{-g}}
\dfrac{\delta S_{\scriptscriptstyle M}}{\delta g^{\mu\nu}}
\end{equation}\\
the equation \eqref{condition3} yields to the Einstein field equations

\begin{equation}\label{Einstein1}
R_{\mu\nu} -
 \dfrac{1}{2} g_{\mu\nu} R =\dfrac{8\pi G}{c^4}T_{\mu\nu} 
 \quad\text{or}\quad G_{\mu\nu}=kT_{\mu\nu} \qquad 
 k=\dfrac{8\pi G}{c^4}
\end{equation}\\
The energy-momentum tensor is introduced as the source term when we want to include any kind of sources, in this case it represents the contribution of matter. The left side of the equation refers to the curvature of spacetime while the right side to the content of it. A reading of these equations in the case of matter is that matter defines the curvature of spacetime while spacetimes curvature dictates the motion of matter.\\

\hspace{-0.65cm} We know that the Einstein tensor satisfies the Bianchi identities \eqref{BianchiEinstein}. So the field equations \eqref{Einstein1} imply that the energy-momentum tensor should also satisfy similar Bianchi identities 

\begin{equation}\label{BianchiEnergy-momentum}
\nabla_{\mu} T^{\mu\nu} = 0 \quad\text{or}\quad 
\nabla^{\mu} T_{\mu\nu} = 0
\end{equation}\\
this implies that locally the energy and the momentum are conservative quantities and it comes with total agreement with Einstein's Principle of Equivalence. \\

\hspace{-0.75cm} However the fact that both the Einstein tensor and the energy-momentum tensor satisfy these 
conservation identities in combination with the metric compatibility \eqref{compatibility} allows us to add an extra term to the field equations

\begin{align}\label{Lambda}
R_{\mu\nu} -
 \dfrac{1}{2} &\ g_{\mu\nu} R  + \Lambda g_{\mu\nu} =
 \dfrac{8\pi G}{c^4}T_{\mu\nu}   \notag\\
  \intertext{\centering or} 
 &\ G_{\mu\nu} + \Lambda g_{\mu\nu}=kT_{\mu\nu}   
 \end{align}\\
Einstein was the first to notice this freedom that his equations provide. He proposed these slightly generalized equations in an attempt to strengthen his theory of a static universe, since the previous equations lead to the conclusion that a static universe can not be stable \cite{Barrow:2003ni}. So
$\Lambda$ was introduced for cosmological purposes and thus it's called \emph{the cosmological constant}. This gave the universe some other dynamics but was still insufficient to justify a static universe and by that time there were also astrological evidence that supported the fact that the universe is not static, most importantly the observations of Hubble that the galaxies are moving away from us supported a model of an expanding universe. This lead Einstein to dismiss his static universe theory and the cosmological constant, famously labeled as his "biggest blunder".\\

\hspace{-0.75cm} We can connect the cosmological constant with the Lagrangian formalism by writing down a corresponding action. This will be nothing more than a constant multiplying the volume element integrated over the entire space. Namely

\begin{equation}
S_{\scriptscriptstyle \Lambda}=
{\lambda}_o \int \sqrt{-g} \; d^4x
\end{equation} \\
So the overall action that leads to the field equations \eqref{Lambda} would be

\begin{equation}\label{SLambda}
S_{tot}= 
S_{\scriptscriptstyle H-E} + 
S_{\scriptscriptstyle \Lambda} +
S_{\scriptscriptstyle M}
\end{equation}\\
and we can easily see that this can take the form

\begin{align}
S_{tot} &=
 \frac{c^4}{16{\pi}G}\int (R-2\Lambda)\sqrt{-g}\;d^4x \; + 
\int \mathcal{L_{\scriptscriptstyle M}} \sqrt{-g}\; d^4x 
\end{align}

\chapter{The FRW Cosmology}

\section{The Cosmological Principle}
After the introduction of GR scientist were able to study the universe in a more mathematical way than ever before. This study of the evolution of the universe, as well as the properies and the dynamics of it is known today as \emph{Cosmology}\cite{Dolgov:2009zj,Plebanski:2006sd,Ryden:2003yy,Uzan:2016wji}. The dynamics of the universe can be described by the Einstein field equations, but to do that we need an appropriate form for the energy-momentum tensor, which is connected to the composition of the universe, and the metric, that is related to the Ricci curvature tensor and the Ricci curvature scalar. So we are called to set up a basis of axioms that we can use in order to construct those objects. This would be \emph{the Cosmological Principle}. 

The Cosmological Principle states that in macroscopic scales the universe can be seen as \emph{homogeneous} and \emph{isotropic}. The \emph{homogeneity} implies that the universe is the same everywhere, meaning that the metric that describes it should be the same in every place of the universe and additionally the curvature too. We can already see that this can not be true in smaller scales like inside a galaxy or our solar system, since we know that massive objects curve the spacetime around them, but on a bigger scope we believe that this holds true and most of the evidence support this as well. 
The \emph{isotropy} means that the universe should look the same across every direction. This suggests that there is no difference in what two different observers in different parts of the universe see. Furthermore it suggests that our own place in the universe is not special by any means. The isotropy of the universe is also heavily supported by astronomical observation, most famously the Cosmic Microwave
Background radiation (CMB)\cite{Durrer:2015lza,White:2002wv}. In fact the CMB appears as an isortopic black body radiation at a temperature of 2.7260±0.0013 K and inhomogeneities of temperature to a factor of $10^{-5}$\cite{Fixsen:2009ug}.

\section{The Robertson-Walker metric}

The metric that is appropriate for a homogeneous and isotropic universe is know as the Robertson-Walker metric \cite{Weinberg:2008zzc,Weinberg:1972kfs,Kolb:1998ai,Liddle:1998ew}

\begin{equation}\label{RW}
ds^2=-dt^2 + a^2(t)\left[\dfrac{dr^2}{1-Kr^2} +
 r^2(d\theta^2 + sin^2\theta \; d\phi^2)\right] 
\end{equation}\\
where we have assumed that c = 1. This metric was firstly introduced by Alexander Friedmann on the year 1922\cite{Friedmann:1924bb,Friedman:1922kd} after solving the Einstein equations under some assumptions for the contents of the universe. The same metric was derived by Howard P. Robertson and  Arthur Geoffrey Walker in the 1930's but in a purely geometrical approach under the assumption of a homogeneous and isotropic universe. Their first approach was purely kinematic and did not predict the function a(t). However, being influenced by Hubble's observations, they introduced this time-only dependent term that can explain the dynamics of the universe as predicted by Hubble's law, hence the function a(t) called the \emph{scale factor}.  \\

\hspace{-0.7cm}The scale factor can be understood by imagining an expansion (or contraction) of the universe according to the following linear rule

\begin{equation}\label{scale factor}
r(t)=a(t)\; r_0
\end{equation}\\
In this relation, r(t) is the distance as measured by an observer at the time \emph{t} .The corresponding coordinate system, which remains the same in time and does not follow the expansion of the universe, is called \emph{physical coordinate system}. Additionaly $r_0$ is the same distance on a coordinate system that follows the expansion (or contraction) of the universe and is called \emph{comoving coordinate system}. In this coordinate system the distance between 2 objects always remains the same during the expansion (or contraction) of the universe.\\

\hspace{-0.7cm}Taking into consideration the existence of the scale factor we can reproduce \\
 Hubble's law 
 
\begin{align}
\dfrac{dr(t)}{dt} &= \dfrac{d}{dt}\big( a(t)\; r_0 \big) \nonumber \\ 
&= \dot a(t)\; r_0 \nonumber \\
&= \dfrac{\dot a(t)}{a(t)} r(t)
\end{align}\\
where we can consider the Hubble parameter to be

\begin{equation}
H \equiv  \dfrac{\dot a(t)}{a(t)}
\end{equation}\\
\\
\\
so we end up with the famous expression of the Hubble's law that describes the expansion of the universe

\begin{equation}
\varv = H\; r
\end{equation}

\hspace{-0.75cm} The geometrical properties of the universe are connected with the constant K  which is associated with the curvature of space. It takes the following 3 values:

\begin{align*}
K = 0 \;\; \Rightarrow &\ \text{space is $R^3$ corresponding to a flat space.} \\
&\ \text{We refer to this case as a \textbf{flat} universe} \\
K = 1 \;\; \Rightarrow &\ \text{space is $S^3$ corresponding to a 3-sphere.}\\
&\ \text{We refer to this case as a \textbf{closed} universe}  \\
K =-1 \Rightarrow &\ \text{space is $H^3$ corresponding to a hyperbolic spatial geometry.}\\
&\ \text{We refer to this case as an \textbf{open} universe} 
\end{align*}
We will later connect the curvature of the universe with the components of it.

\section{The Friedmann equations}

Writing down the elements of the Robertson-Walker metric we have
\begin{align}
\begin{split}\label{FRW metric}
&\ g_{tt} \equiv  g_{00} = -1 \\
&\ g_{rr} \equiv  g_{11} = \dfrac{a^2(t)}{1-Kr^2} \\
&\ g_{\theta \theta} \equiv g_{22} = a^2(t)r^2 \\
&\ g_{\phi \phi}  \equiv g_{33} = a^2(t)r^2\; sin^2\theta \\
&\ g_{\mu\nu} = 0 \quad\text{for}\quad \mu \neq \nu
\end{split}
\end{align}\\
and for the inverse metric we can easily get from 
\eqref{invmetric} 

\begin{align}
\begin{split}\label{FRW invmetric}
&\ g^{tt} \equiv  g^{00} = -1 \\
&\ g^{rr} \equiv  g^{11} = \dfrac{1-Kr^2}{a^2(t)} \\
&\ g^{\theta \theta} \equiv g^{22} = \dfrac{1}{a^2(t)r^2} \\
&\ g^{\phi \phi}  \equiv g^{33} =\dfrac{1}{a^2(t)r^2\; sin^2\theta} \\
&\ g^{\mu\nu} = 0 \quad\text{for}\quad \mu \neq \nu
\end{split}
\end{align}\\
In addition the partial derivatives of the metric are

\begin{align}\label{metricderivatives}
\begin{split}
\partial_\mu g_{ij} &= 0 \quad\forall \mu,  i\neq j \\
\partial_\mu g_{00} &= 0 \quad\forall \mu \\
\partial_0 g_{11} &=\dfrac{\partial}{\partial t}\left(\dfrac{a^2(t)}{1-Kr^2}\right)
= \dfrac{2 \dot a a}{1-Kr^2} \\
 \partial_1 g_{11} &=\dfrac{\partial}{\partial r}\left(\dfrac{a^2(t)}{1-Kr^2}\right)
= \dfrac{2Kr\; a^2}{(1-Kr^2)^2} \\
\partial_2 g_{11} &=\dfrac{\partial}{\partial \theta}\left(\dfrac{a^2(t)}{1-Kr^2}\right) = 0 \\
\partial_3 g_{11} &=\dfrac{\partial}{\partial \phi}\left(\dfrac{a^2(t)}{1-Kr^2}\right) = 0 \\
\partial_0 g_{22} &= 2\dot aar^2 \\
\partial_1 g_{22} &= 2 a^2 r \\
\partial_2 g_{22} &= \partial_3 g_{22} = 0 \\
\partial_0 g_{33} &= 2\dot a ar^2sin^2\theta \\
\partial_1 g_{33} &= 2a^2rsin^2\theta	\\
\partial_2 g_{33} &= 2a^2r^2sin\theta cos\theta \\
\partial_3 g_{33} &= 0
\end{split}
\end{align}
We can now use the formula \eqref{Chrisoffel} by substituting the expressions from \eqref{FRW invmetric} and \eqref{metricderivatives} we get in order to compute the Christoffel symbols. Assuming the symmetry of the Christoffel symbols in regard to the lower indices

\begin{equation}
\Gamma^{\alpha}_{\mu\nu} = \Gamma^{\alpha}_{\nu\mu}
\end{equation}\\
 we expect 40 different ones, 10 for each value of $a$. However in the case of the Robertson-Walker metric the non-trivial ones are way less. The computation is pretty straightforward. For example, for a non-zero one
 
\begin{align*}
\Gamma^{0}_{11} &= \frac{1}{2} g^{0k} \big(
\partial_1 g_{k1} + \partial_1 g_{1k} - \partial_k g_{11} \big) \\
&= \frac{1}{2} g^{00} \big(
\partial_1 g_{01} + \partial_1 g_{10} - \partial_0 g_{11} \big) \\ 
&\ \;  + \frac{1}{2} g^{01} \big(
\partial_1 g_{11} + \partial_1 g_{11} - \partial_1 g_{11} \big) \\ 
&\ \; + \frac{1}{2} g^{02} \big(
\partial_1 g_{21} + \partial_1 g_{12} - \partial_2 g_{11} \big) \\ 
&\ \;+ \frac{1}{2} g^{03} \big(
\partial_1 g_{31} + \partial_1 g_{13} - \partial_3 g_{11} \big) \\
&=\frac{1}{2} g^{00} \big(-\partial_0 g_{11} \big) 
= \frac{1}{2}(-1) \left(-\dfrac{2\dot a a}{1-Kr^2}\right) 
= \dfrac{\dot a a}{1-Kr^2}
\end{align*}\\
In a similar we compute the remaining Christoffel symbols. We end up with:
\begin{align}\label{FRW Christoffel}
&\ 1) \;\;\Gamma^{0}_{00}=0 \\ \nonumber
&\ 2) \;\;\Gamma^{0}_{01}=\Gamma^{0}_{10}=0\\ \nonumber
&\ 3) \;\;\Gamma^{0}_{02}=\Gamma^{0}_{20}=0\\ \nonumber
&\ 4) \;\;\Gamma^{0}_{03}=\Gamma^{0}_{30}=0\\\nonumber
&\ 5) \;\;\Gamma^{0}_{11}=\dfrac{\dot a a}{1-Kr^2} \\ \nonumber
&\ 6) \;\;\Gamma^{0}_{12} = \Gamma^{0}_{21} = 0\\ \nonumber
&\ 7)\;\; \Gamma^{0}_{13} = \Gamma^{0}_{31} =0\\ \nonumber
&\ 8) \;\;\Gamma^{0}_{22} = \dot a ar^2\\ \nonumber
&\ 9) \;\;\Gamma^{0}_{23} = \Gamma^{0}_{32} =0\\ \nonumber
&\ 10) \;\;\Gamma^{0}_{33} = \dot aar^2sin^2\theta \\ \nonumber
&\ 11) \;\;\Gamma^{1}_{00} =0\\ \nonumber
&\ 12) \;\;\Gamma^{1}_{01} = \Gamma^{1}_{10} = \dfrac{\dot a}{a}\\ \nonumber
&\ 13) \;\;\Gamma^{1}_{02} = \Gamma^{1}_{20} = 0 \\ \nonumber
&\ 14) \;\;\Gamma^{1}_{03} = \Gamma^{1}_{30} = 0\\ \nonumber
&\ 15)\;\; \Gamma^{1}_{11} = \dfrac{Kr}{1-Kr^2} \\ \nonumber
&\ 16) \;\;\Gamma^{1}_{12} = \Gamma^{1}_{21} = 0\\ \nonumber
&\ 17) \;\;\Gamma^{1}_{13} = \Gamma^{1}_{31}=0\\ \nonumber
&\ 18) \;\;\Gamma^{1}_{22} = -r(1-Kr^2)\\ \nonumber
&\ 19) \;\;\Gamma^{1}_{23} = \Gamma^{1}_{32} = 0\\ \nonumber
&\ 20) \;\;\Gamma^{1}_{33} = -r(1-Kr^2)sin^2\theta\\ \nonumber
&\ 21) \;\;\Gamma^{2}_{00}=0\\ \nonumber
&\ 22) \;\;\Gamma^{2}_{01}=\Gamma^{2}_{10}=0\\ \nonumber
&\ 23) \;\;\Gamma^{2}_{02}=\Gamma^{2}_{20}= \dfrac{\dot a}{a}\\ \nonumber
&\ 24) \;\;\Gamma^{2}_{03}=\Gamma^{2}_{30}=0\\ \nonumber
&\ 25)\;\; \Gamma^{2}_{11}=0\\ \nonumber
&\ 26)\;\; \Gamma^{2}_{12}=\Gamma^{2}_{21} = \frac{1}{r}\\ \nonumber
&\ 27) \;\;\Gamma^{2}_{13}=\Gamma^{2}_{31}=0\\ \nonumber
&\ 28) \;\;\Gamma^{2}_{22}=0\\ \nonumber
&\ 29) \;\;\Gamma^{2}_{23}=\Gamma^{2}_{32}=0\\ \nonumber
&\ 30) \;\;\Gamma^{2}_{33}=-sin\theta \; cos\theta  \\ \nonumber
&\ 31)\;\; \Gamma^{3}_{00}=0\\ \nonumber
&\ 32)\;\; \Gamma^{3}_{01}=\Gamma^{3}_{10}=0\\ \nonumber
&\ 33) \;\;\Gamma^{3}_{02}=\Gamma^{3}_{30}=0\\ \nonumber
&\ 34)\;\; \Gamma^{3}_{03}=\Gamma^{3}_{30}=\dfrac{\dot a}{a}\\ \nonumber
&\ 35)\;\; \Gamma^{3}_{11}=0\\ \nonumber
&\ 36) \;\;\Gamma^{3}_{12}=\Gamma^{3}_{21}= 0\\ \nonumber
&\ 37) \;\;\Gamma^{3}_{13}=\Gamma^{3}_{31}=\frac{1}{r}\\ \nonumber
&\ 38)\;\; \Gamma^{3}_{22}=0\\ \nonumber
&\ 39)\;\; \Gamma^{3}_{23}=\Gamma^{3}_{32}=\dfrac{cos\theta}{sin\theta}=cot\theta \\ \nonumber
&\ 40)\;\; \Gamma^{3}_{33}= 0  \nonumber
\end{align}
\\
\hspace{-0.7cm}Next comes the computation of the Ricci curvature tensor from the formula \eqref{Ricci}, as we have the expressions for the Christoffel symbols. First, we are going to show that $R_{\mu\nu}=0 \quad\forall \mu \neq \nu$. For that we calculate the terms $ \Gamma^\kappa_{\;\mu\kappa}$ 

\begin{align*}
\Gamma^\kappa_{\;0\kappa} &=
\Gamma^0_{\;00}+
\Gamma^1_{\;01}+
\Gamma^2_{\;02}+
\Gamma^3_{\;03} =
3\dfrac{\dot a}{a}\\
\Gamma^\kappa_{\;1\kappa} &=
\Gamma^0_{\;10}+
\Gamma^1_{\;11}+
\Gamma^2_{\;12}+
\Gamma^3_{\;13} = \dfrac{Kr}{1-Kr^2} + \frac{2}{r}\\
\Gamma^\kappa_{\;2\kappa} &=
\Gamma^0_{\;20}+
\Gamma^1_{\;21}+
\Gamma^2_{\;22}+
\Gamma^3_{\;23} = cot\theta \\
\Gamma^\kappa_{\;3\kappa} &=
\Gamma^0_{\;30}+
\Gamma^1_{\;31}+
\Gamma^2_{\;32}+
\Gamma^3_{\;33} = 0
\end{align*}\\
\\
we notice that the first expression is a function of just the time $t$, the second one of just the distance $r$, the third one of just $\theta$ while the last one is zero. So for the case of $\mu \neq \nu$ it is actually $\partial_\nu  \Gamma^\kappa_{\;\mu\kappa}=0$ since

\begin{align*}
\partial_0  \Gamma^\kappa_{\;\mu\kappa} &= 0 
\qquad\text{$\forall \mu \neq 0$ }  \\
\partial_1  \Gamma^\kappa_{\;\mu\kappa} &= 0 
\qquad\text{$\forall \mu \neq 1$ }  \\
\partial_2  \Gamma^\kappa_{\;\mu\kappa} &= 0 
\qquad\text{$\forall \mu \neq 2$ }  \\
\partial_3  \Gamma^\kappa_{\;\mu\kappa} &= 0 
\end{align*}\\
Moreover the non-zero Christoffel symbols for   $\mu \neq \nu$ are

\begin{align*}
\Gamma^1_{\; 01} &=
 \Gamma^2_{\; 02} =
 \Gamma^3_{\; 03} =
 \dfrac{\dot a}{a} \\
 \Gamma^2_{\; 12} &=
  \Gamma^3_{\; 13} =
  \frac{1}{r} \\
  \Gamma^3_{\; 23} &= cot\theta
\end{align*}\\
so it's easy to see that the second partial derivative of the Christoffel symbols that appears on the formula of the Ricci tensor turns out to be zero as well

\begin{align*}
\partial_\kappa \Gamma^\kappa_{\;\mu\nu}=0 
\quad\text{$\forall \mu \neq \nu$}
\end{align*}\\
The expression for the Ricci tensor now becomes

\begin{equation}
R_{\mu\nu} = 
\Gamma^{\lambda}_{\;\nu\mu}\Gamma^{\kappa}_{\;\kappa\lambda} - \Gamma^{\lambda}_{\;\kappa\mu}\Gamma^{\kappa}_{\;\nu\lambda}
\qquad\text{$ \mu \neq \nu$}
\end{equation}  \\
By expanding the first term we get

\begin{align}
\Gamma^{\lambda}_{\;\nu\mu}\Gamma^{\kappa}_{\;\kappa\lambda} =
\Gamma^{0}_{\;\nu\mu}\Gamma^{\kappa}_{\;\kappa 0} +
\Gamma^{1}_{\;\nu\mu}\Gamma^{\kappa}_{\;\kappa 1} +
\Gamma^{2}_{\;\nu\mu}\Gamma^{\kappa}_{\;\kappa 2} +
\Gamma^{3}_{\;\nu\mu}\Gamma^{\kappa}_{\;\kappa 3}
\end{align}\\
where $\Gamma^{\kappa}_{\;\kappa 3} = 0$ , as we calculated before, and $\Gamma^{0}_{\;\nu\mu}=0 \quad\forall \mu \neq \nu$. \\
So, by keeping only the non-zero Christoffel symbols , the first term becomes

\begin{align}\label{term1}
\begin{split}
\Gamma^{\lambda}_{\;\nu\mu}\Gamma^{\kappa}_{\;\kappa\lambda} &=
\Gamma^{1}_{\;\nu\mu}\Gamma^{\kappa}_{\;\kappa 1} +
\Gamma^{2}_{\;\nu\mu}\Gamma^{\kappa}_{\;\kappa 2} \\
&= \Gamma^{1}_{\; 01}\Gamma^{\kappa}_{\;\kappa 1} +
\Gamma^{2}_{\; 02}\Gamma^{\kappa}_{\;\kappa 2} +
\Gamma^{2}_{\; 12}\Gamma^{\kappa}_{\;\kappa 2}
\end{split}
\end{align}\\
In a similar way we expand the second term

\begin{align*}
\Gamma^{\lambda}_{\;\kappa\mu}\Gamma^{\kappa}_{\;\nu\lambda} &= \;\;
\Gamma^{0}_{\;\kappa\mu}\Gamma^{\kappa}_{\;\nu 0} +
\Gamma^{1}_{\;\kappa\mu}\Gamma^{\kappa}_{\;\nu 1} + 
\Gamma^{2}_{\;\kappa\mu}\Gamma^{\kappa}_{\;\nu 2} +\Gamma^{3}_{\;\kappa\mu}\Gamma^{\kappa}_{\;\nu 3}  \\
&= \;\;  \Gamma^{0}_{\; 0 \mu}\Gamma^{0}_{\;\nu 0} +
\Gamma^{1}_{\; 0 \mu}\Gamma^{0}_{\;\nu 1} + 
\Gamma^{2}_{\; 0 \mu}\Gamma^{0}_{\;\nu 2} +
\Gamma^{3}_{\; 0 \mu}\Gamma^{0}_{\;\nu 3}  \\
   &\ \;\; + 
\Gamma^{0}_{\; 1 \mu}\Gamma^{1}_{\;\nu 0} +
\Gamma^{1}_{\; 1 \mu}\Gamma^{1}_{\;\nu 1} + 
\Gamma^{2}_{\; 1 \mu}\Gamma^{1}_{\;\nu 2} +
\Gamma^{3}_{\; 1 \mu}\Gamma^{1}_{\;\nu 3}  \\
&\ \;\; +
\Gamma^{0}_{\; 2 \mu}\Gamma^{2}_{\;\nu 0} +
\Gamma^{1}_{\; 2 \mu}\Gamma^{2}_{\;\nu 1} + 
\Gamma^{2}_{\; 2 \mu}\Gamma^{2}_{\;\nu 2} +
\Gamma^{3}_{\; 2 \mu}\Gamma^{2}_{\;\nu 3}  \\
&\ \;\; +
\Gamma^{0}_{\; 3 \mu}\Gamma^{3}_{\;\nu 0} +
\Gamma^{1}_{\; 3 \mu}\Gamma^{3}_{\;\nu 1} + 
\Gamma^{2}_{\; 3 \mu}\Gamma^{3}_{\;\nu 2} +
\Gamma^{3}_{\; 3 \mu}\Gamma^{3}_{\;\nu 3}  \\
\end{align*}\\
and for $\mu \neq \nu$ the only the non-zero terms are

\begin{align}\label{term2}
\Gamma^{\lambda}_{\;\kappa\mu}\Gamma^{\kappa}_{\;\nu\lambda} =
\Gamma^1_{10} \Gamma^1_{11} +
\Gamma^2_{20} \Gamma^2_{12} +
\Gamma^3_{3 \mu} \Gamma^3_{3 \nu}
\end{align}\\
In order to compare the expressions \eqref{term1} and \eqref{term2} we need to consider different combinations for $\mu$ and $ \nu$

\begin{align*}
\text{for \; $\mu ,\nu =(1,2,3)$}: \qquad
&\ \Gamma^{\lambda}_{\;\nu\mu}\Gamma^{\kappa}_{\;\kappa\lambda} =
\Gamma^{2}_{12}\Gamma^{\kappa}_{\;2 \kappa} =
\frac{1}{r} cot\theta \\
\\
&\ \Gamma^{\lambda}_{\;\kappa\mu}\Gamma^{\kappa}_{\;\nu\lambda} = 
\Gamma^3_{31} \Gamma^3_{32} =
\frac{1}{r} cot\theta \\
\\
\\
\
\text{for \; $\mu =0 ,\nu =(1,3)$}: \quad
&\ \Gamma^{\lambda}_{\;\nu\mu}\Gamma^{\kappa}_{\;\kappa\lambda} =
\Gamma^{1}_{\; 01}\Gamma^{\kappa}_{\;\kappa 1} =
\dfrac{\dot a}{a}\left(\dfrac{Kr}{1-Kr^2}+\frac{2}{r}\right)\\
\\
&\ \Gamma^{\lambda}_{\;\kappa\mu}\Gamma^{\kappa}_{\;\nu\lambda} =
\Gamma^1_{10} \Gamma^1_{11} +
\Gamma^2_{20} \Gamma^2_{12} +
\Gamma^3_{3 0} \Gamma^3_{3 1}=
\dfrac{\dot a}{a}\left(\dfrac{Kr}{1-Kr^2}+\frac{2}{r}\right)\\
\\
\\
\text{for \; $\mu =0,\nu =2$}: \qquad\;\;
&\ \Gamma^{\lambda}_{\;\nu\mu}\Gamma^{\kappa}_{\;\kappa\lambda} =
\Gamma^{2}_{\; 02}\Gamma^{\kappa}_{\;\kappa 2} =
\dfrac{\dot a}{a}cot\theta \\
\\
&\ \Gamma^{\lambda}_{\;\kappa\mu}\Gamma^{\kappa}_{\;\nu\lambda} = 
\Gamma^3_{3 0} \Gamma^3_{3 2} =
\dfrac{\dot a}{a}cot\theta
\end{align*}\\
Those are all the possible combinations for $\mu$ and $\nu$ so we can conclude that those two expressions are indeed equal in the case of $\mu \neq \nu$. Thus we can claim that for the case of the Robertson-Walker metric

\begin{equation}
R_{\mu\nu} = 0 \quad\text{$\forall \mu \neq \nu$}
\end{equation}
This means that the only non-zero elements of the Ricci curvature tensor are only 4; $R_{00}$, $R_{11}$, 
$R_{22}$, $R_{33}$.

\begin{align*}
R_{00}=
\partial_\kappa \Gamma^\kappa_{\;00} -
\partial_0 \Gamma^\kappa_{\;0 \kappa}+
\Gamma^\lambda_{\;00} \Gamma^\kappa_{\;\lambda \kappa} -
\Gamma^\lambda_{\;\kappa 0}\Gamma^\kappa_{\;0\lambda}
\end{align*}\\
where

\begin{align*}
&\ \Gamma^\kappa_{\;00}=0 \\
&\ \Gamma^\kappa_{\;0\kappa}=3\dfrac{\dot a}{a}\\
&\ \Gamma^\lambda_{\;\kappa 0}\Gamma^\kappa_{\;0\lambda}=
\left(\Gamma^1_{01}\right)^2+
\left(\Gamma^2_{02}\right)^2+
\left(\Gamma^3_{03}\right)^2 =
\left(\dfrac{\dot a}{a}\right)^2+
\left(\dfrac{\dot a}{a}\right)^2+
\left(\dfrac{\dot a}{a}\right)^2 = 
3\left(\dfrac{\dot a}{a}\right)^2
\end{align*}\\
so for $R_{00}$ we end up with

\begin{align*}
R_{00}&= 
\partial_0 \left(3\dfrac{\dot a}{a}\right)-
3\left(\dfrac{\dot a}{a}\right)^2 \\
&=-3\dfrac{\ddot a}{a} - 
3\dfrac{\dot a^2}{a^2}+
3\left(\dfrac{\dot a}{a}\right)^2 \\
&= -3\dfrac{\ddot a}{a} \\
\end{align*}\\
For the $R_{11}$ element

\begin{align*}
R_{11}&= \;\;
\partial_\kappa \Gamma^\kappa_{\;11} -
\partial_1 \Gamma^\kappa_{\;1 \kappa}+
\Gamma^\lambda_{\;11} \Gamma^\kappa_{\;\lambda \kappa} -
\Gamma^\lambda_{\;\kappa 1}\Gamma^\kappa_{\;1\lambda} \\ \\
&=\;\; 
\partial_0 \Gamma^0_{\;11} -
\partial_1 \Gamma^\kappa_{\;1 \kappa}+
\Gamma^0_{\;11} \Gamma^\kappa_{\;0 \kappa} -
\Gamma^0_{\;\kappa 1}\Gamma^\kappa_{\;1 0} \\
&\ \;\; +
\partial_1 \Gamma^1_{\;11} 		
\qquad \quad\;\; +
\Gamma^1_{\;11} \Gamma^\kappa_{\;1 \kappa} -
\Gamma^1_{\;\kappa 1}\Gamma^\kappa_{\;1 1} \\
&\ \;\; +
\partial_2 \Gamma^2_{\;11} 		
\qquad \quad\;\; +
\Gamma^2_{\;11} \Gamma^\kappa_{\;2  \kappa} -
\Gamma^2_{\;\kappa 1}\Gamma^\kappa_{\;1 2} \\
&\ \;\; +
\partial_3 \Gamma^3_{\;11} 		
\qquad \quad\;\; +
\Gamma^3_{\;11} \Gamma^\kappa_{\; 3 \kappa} -
\Gamma^3_{\;\kappa 1}\Gamma^\kappa_{\;1 3} 
\end{align*}\\
and by keeping only the non-zero terms we get

\begin{align*}
R_{11}&=\;\; 
\partial_0 \Gamma^0_{\;11} -
\partial_1 \Gamma^\kappa_{\;1 \kappa}+
\Gamma^0_{\;11} \Gamma^\kappa_{\;0 \kappa} -
\Gamma^0_{\;\kappa 1}\Gamma^\kappa_{\;1 0} \\
&\ \;\; +
\partial_1 \Gamma^1_{\;11} 		
\qquad \quad\;\; +
\Gamma^1_{\;11} \Gamma^\kappa_{\;1 \kappa} -
\Gamma^1_{\;\kappa 1}\Gamma^\kappa_{\;1 1} \\
&\ \;\; 
\qquad \quad\;\; 		
\qquad \qquad\;\; 
\qquad \quad\;\; -
\Gamma^2_{\;\kappa 1}\Gamma^\kappa_{\;1 2} \\
&\ \;\; 
\qquad \quad\;\;	
\qquad \qquad\;\; 
\qquad \quad\;\; -
\Gamma^3_{\;\kappa 1}\Gamma^\kappa_{\;1 3} 
\end{align*}\\
and further expanding whilst ignoring the non-zero terms

\begin{align*}
R_{11}&=\;\;
\partial_0 \Gamma^0_{\;11} -
\partial_1 \Gamma^\kappa_{\;1 \kappa}+
\Gamma^0_{\;11} \Gamma^\kappa_{\;0 \kappa} -
\Gamma^0_{\;0 1}\Gamma^0_{\;1 0} \\
&\ \;\; +
\partial_1 \Gamma^1_{\;11} 		
\qquad \quad\;\; +
\Gamma^1_{\;11} \Gamma^\kappa_{\;1 \kappa} -
\Gamma^1_{\;0 1}\Gamma^0_{\;1 1} -
\Gamma^1_{\;1 1}\Gamma^1_{\;1 1} \\
&\ \;\; 
\qquad \quad\;\; 		
\qquad \qquad\;\; 
\qquad \quad\;\; -
\Gamma^2_{\;2 1}\Gamma^2_{\;1 2} \\
&\ \;\; 
\qquad \quad\;\;	
\qquad \qquad\;\; 
\qquad \quad\;\; -
\Gamma^3_{\;2 1}\Gamma^2_{\;1 3} \\
\end{align*}\\
\\
\\
\\
\\
where substituting the expressions for the Christoffel symbols gives
\\
\begin{align*}
R_{11}&=\;\;
\partial_0 \left(\dfrac{\dot a a}{1-Kr^2}\right) -
\partial_1 \left(\dfrac{Kr}{1-Kr^2}+\frac{2}{r}\right) +
\left(\dfrac{\dot a a}{1-Kr^2}\right)\left(3\dfrac{\dot a}{a}\right) -
\left(\dfrac{\dot a a}{1-Kr^2}\right)\left(\dfrac{\dot a}{a}\right)  \\
&\ \;\; + \partial_1\left(\dfrac{Kr}{1-Kr^2}\right)
\qquad\qquad +
\left(\dfrac{Kr}{1-Kr^2}\right)\left(\dfrac{Kr}{1-Kr^2} +\frac{2}{r}\right) -
\left(\dfrac{\dot a a}{1-Kr^2}\right)\left(\dfrac{\dot a}{a}\right) -
\left(\dfrac{Kr}{1-Kr^2}\right)^2\\
&\ \hspace{9.5cm}- \frac{1}{r^2} \\
&\ \hspace{9.5cm}- \frac{1}{r^2} \\
\\
&= \dfrac{\ddot a a}{1-Kr^2} +
\dfrac{\dot a^2}{1-Kr^2} +
\dfrac{\dot a^2}{1-Kr^2} +
\dfrac{Kr}{1-Kr^2}\left(\frac{2}{r}\right) \\
&= \frac{1}{1-Kr^2}\left(\ddot a a + 2\dot a^2 +2K\right) \\
&= \frac{a^2}{1-Kr^2}\left( \dfrac{\ddot a}{a} 
+ 2\left(\dfrac{\dot a}{a}\right)^2 +
2\dfrac{K}{a^2}\right)
\end{align*}\\
we work in a similar way for the remaining 2 elements of the Ricci tensor and we end up with

\begin{align}\label{FRW Ricci1}
\begin{split}
&\ R_{00}=-3\dfrac{\ddot a}{a}=3g_{00}\left(\dfrac{\ddot a}{a}\right)\\
\\
&\ R_{11}=
   \frac{a^2}{1-Kr^2}\left( \dfrac{\ddot a}{a} 
   + 2\left(\dfrac{\dot a}{a}\right)^2 +
   2\dfrac{K}{a^2}\right) =
   g_{11}\left( \dfrac{\ddot a}{a} 
   + 2\left(\dfrac{\dot a}{a}\right)^2 +
   2\dfrac{K}{a^2}\right) \\
   \\
&\ R_{22}= \;\;\;
   a^2 \; r^2 \;\;\left( \dfrac{\ddot a}{a} 
   + 2\left(\dfrac{\dot a}{a}\right)^2 +
   2\dfrac{K}{a^2}\right) =
   g_{22}\left( \dfrac{\ddot a}{a} 
   + 2\left(\dfrac{\dot a}{a}\right)^2 +
   2\dfrac{K}{a^2}\right) \\
   \\
&\ R_{33}=
   a^2  r^2sin^2\theta \left( \dfrac{\ddot a}{a} 
   + 2\left(\dfrac{\dot a}{a}\right)^2 +
   2\dfrac{K}{a^2}\right) =
   g_{33}\left( \dfrac{\ddot a}{a} 
   + 2\left(\dfrac{\dot a}{a}\right)^2 +
   2\dfrac{K}{a^2}\right) \\   	
\end{split} 
\end{align}\\
It is easy now to calculate the Ricci scalar by contracting the Ricci tensor \eqref{Ricciscalar}

\begin{align*}
R&=g^{\mu\nu} R_{\mu\nu} = R^\mu_{\;\mu} =
   R^0_{\;0} +
   R^1_{\;1}+
   R^2_{\;2}+
   R^3_{\;3}
\end{align*}\\
where

\begin{align}\label{FRW Ricci2}
\begin{split}
&\ R^0_{\;0} = 
g^{00}R_{00}=g^{00}g_{00}\; 3\left(\dfrac{\ddot a}{a}\right) =
3\dfrac{\ddot a}{a} \\
\\
&\ R^1_{\;1} =
g^{11}R_{11} = g^{11}g_{11}
   \left( \dfrac{\ddot a}{a} 
   + 2\left(\dfrac{\dot a}{a}\right)^2 +
   2\dfrac{K}{a^2}\right) =
   \left( \dfrac{\ddot a}{a} 
   + 2\left(\dfrac{\dot a}{a}\right)^2 +
   2\dfrac{K}{a^2}\right)\\
\\
&\ R^2_{\;2} =
   g^{22}R_{22} = g^{22}g_{22}
   \left( \dfrac{\ddot a}{a} 
   + 2\left(\dfrac{\dot a}{a}\right)^2 +
   2\dfrac{K}{a^2}\right) =
   \left( \dfrac{\ddot a}{a} 
   + 2\left(\dfrac{\dot a}{a}\right)^2 +
   2\dfrac{K}{a^2}\right)  \\
\\   
&\ R^3_{\;3} =
   g^{33}R_{33} = g^{33}g_{33}
   \left( \dfrac{\ddot a}{a} 
   + 2\left(\dfrac{\dot a}{a}\right)^2 +
   2\dfrac{K}{a^2}\right) =
   \left( \dfrac{\ddot a}{a} 
   + 2\left(\dfrac{\dot a}{a}\right)^2 +
   2\dfrac{K}{a^2}\right)   
\end{split}
\end{align}\\
And we end up with the following value for the Ricci scalar

\begin{equation}\label{FRW Ricciscalar}
R= 6\left( \dfrac{\ddot a}{a} 
   + \left(\dfrac{\dot a}{a}\right)^2 +
   \dfrac{K}{a^2}\right)
\end{equation}\\
\\
\\
\\
\\
\\
We can also calculate the values for the Einstein tensor \eqref{EinsteinTensor}

\begin{align}\label{FRW Einstein}
\begin{split}
&\ G^0_{\; 0} = 
-\left(3\left(\dfrac{\dot a}{a}\right)^2 +
3\dfrac{K}{a^2}\right) \\
\\
&\ G^1_{\; 1}=G^2_{\; 2}=G^3_{\; 3} =
-  \left( 2\dfrac{\ddot a}{a} 
   + \left(\dfrac{\dot a}{a}\right)^2 +
   \dfrac{K}{a^2}\right)
\end{split}
\end{align}\\
This gives us everything we need to calculate the left side of the Einstein field equations \eqref{Einstein1}. The number of the Einstein equations has also dropped down to 2, due to the symmetries that the Robertson-Walker metric provides, instead of 10, given that the Einstein tensor is symmetric. \\

\hspace{-0.7cm} The only thing that remains is to get a proper form for the energy-momentum tensor. This comes out of the assumption that the universe behaves like a perfect fluid. For a perfect fluid we know that the energy-momentum tensor takes the following form

\begin{equation}\label{Tmn1}
T_{\mu\nu}= (p+\rho)U_\mu U_\nu + pg_{\mu\nu}
\end{equation}\\
where we have assumed c=1. The term $U_\mu$ is the 4-velocity and additionally $\rho$, p the density and pressure of the fluid respectively, which are functions of just time. In cosmology
the density refers to the density of matter in the universe and the pressure to the relativistic particles. We usually 
refer to the pressure term as the radiation pressure.\\

\hspace{-0.7cm}Since we have assumed an isotropic and homogeneous universe, the model of a perfect fluid gives us an isotropic and homogeneous fluid as well. Thus we can elect to express the energy-momentum tensor in the frame of the comoving reference system. in that case the 4-velocity takes the following simple from

\begin{align*}
U_\mu = (-1,0,0,0)
\end{align*}\\
The only non-zero component of the 4-velocity is the time component. That is because in the comoving reference system any body is at rest with the fluid, or the universe in this case, so the spatial components of the velocity must be zero. We also know that the 4-velocity satisfies the rule

\begin{align*}
U_\mu U^\mu = -1
\end{align*}\\
so that justifies the form of the 4-velocity in this case. We then end up with the followin form for the energy-momentum tensor

\begin{align*}
T^\mu_{\;\nu} = g^{\mu\lambda}T_{\lambda\nu} = diag(-\rho,p,p,p)
\end{align*}\\
Now we have everything required to write down the Einstein equations for this case of the Robertson-Walker metric. As stated those came down to being just 2, one representing the time component and the other the spatial one. For the time component we get

\begin{equation}
G^0_{\;0}=8\pi G \; T^0_{\;0 }
\end{equation} \\
\\
\\
\\
which leads to the following equation

\begin{equation}\label{Friedmann1}
\left(\dfrac{\dot a}{a}\right)^2 =
\dfrac{8\pi G}{3}\rho - \dfrac{K}{a^2}
\end{equation}\\
In a similar way the spatial component of the Einstein tensor leads to the equation

\begin{equation}\label{Friedmann2}
2\dfrac{\ddot a}{a} +
\left(\dfrac{\dot a}{a}\right)^2 + 
\dfrac{K}{a^2} =
-8\pi Gp
\end{equation}\\
we can rewrite those equations using the Hubble parameter H and its time \\ derivative

\begin{align}
\dot H = 
\dfrac{d}{dt}\left(\dfrac{\dot a}{a}\right) =
\dfrac{\ddot a}{a} - \left(\dfrac{\dot a}{a}\right)^2
\end{align}\\
Then we end up with

\begin{align}
&\ H^2 =
\dfrac{8\pi G}{3}\rho - \dfrac{K}{a^2} \label{Friedmann3}\\ \nonumber
\\ 
&\ \dot H = - 4\pi G (\rho + p) +
\dfrac{K}{a^2}\label{Friedmann4}
\end{align}\\
The above equations are called the \emph{the Friedmann equations} and a universe that is described by the Robertson-Walker metric and those equations is called Friedmann-Robertson-Walker universe or FRW universe in short.\\
\\
\\
Furthermore by subtracting the equations \eqref{Friedmann1} and \eqref{Friedmann2} we get

\begin{equation}\label{acceleration equation}
\dfrac{\ddot a}{a}= -\dfrac{4\pi G}{3}(\rho +3p)
\end{equation}\\
This equation is called \emph{the acceleration equation} since it connects the expansion rate of the universe with just the components of it. The scale factor $a(t)$ is inherently a positive value, so the sign of\; $\ddot a$\; depends only from the term $(\rho + 3p)$. \\

\hspace{-0.7cm} We can also combine the Friedmann equation in another way by taking the partial derivative of the first \eqref{Friedmann2} , which gives

\begin{align*}
2H\dot H - \dfrac{8 \pi G}{3}\dot \rho =
2K \dfrac{\dot a}{a^3}
\end{align*}\\
and then substituting the $K$ from the second equation \eqref{Friedmann4}, which gives 

\begin{align*}
K= a^2\dot H + 4\pi G\; a^2(\rho + p)
\end{align*}\\
Then with some simple algebra we derive the following equation

\begin{equation}\label{continuity}
\dot \rho +3H(\rho + p)= 0
\end{equation}\\
which is named \emph{continuity equation} as it is the equivalent of the ordinary continuity equation when we apply it on the universe. To further understand why this equation corresponds to a conservation law we are going to derive this from the conservation of the energy-momentum tensor, which is described via the Bianchi identities \eqref{BianchiEnergy-momentum} and translates to a generalized form of the conservation of energy and momentum (or mass).

\begin{align*}
\nabla_\mu T^\mu_{\; \nu}= 0 \Rightarrow
\partial_\mu  T^\mu_{\; \nu} -
\Gamma^\alpha_{\; \mu \nu}T^\mu_{\; \alpha} +
\Gamma^\mu_{\; \alpha \mu}T^\alpha_{\; \nu}=0 
\end{align*}\\
For simplicity we are going to consider the case of a flat FRW universe that is described by the metric

\begin{equation}\label{FRW flat}
g_{\mu\nu}=diag(-1,a^2,a^2,a^2)
\end{equation}\\
 Then the equation for $\nu=0$ leads us to
 
\begin{align*}
&\ \partial_\mu  T^\mu_{\; 0} = -\dot \rho \\
&\ \Gamma^\alpha_{\; \mu 0}=
\frac{1}{2}g^{\alpha \kappa}
\left( \partial_\mu g_{\kappa 0} + 
\partial_0 g_{\mu \kappa} -
\partial_\kappa g_{\mu 0}\right)\\
&\ \Gamma^\mu_{\; \alpha \mu}T^\alpha_{\; 0} = 
\Gamma^\mu_{\; 0 \mu}T^0_{\; 0}
\end{align*}\\
where for $\mu=0$ we can see that $\Gamma^\alpha_{\; \mu 0} = 0
= \Gamma^0_{\; 0 0}$. So for $\mu \neq 0$ we calculate the expressions

\begin{align*}
\Gamma^\mu_{\; \mu 0} &=
\frac{1}{2}g^{\mu \kappa}
\left( \partial_\mu g_{\kappa 0} + 
\partial_0 g_{\mu \kappa} -
\partial_\kappa g_{\mu 0}\right) \\
&= \frac{1}{2} g^{\mu \kappa} (\partial_0 g_{\mu \kappa}) \\ 
&= \frac{1}{2} ( g^{11} \partial_0 g_{11}+
g^{22} \partial_0 g_{22}+
g^{33} \partial_0 g_{33}) \\
&= \frac{3}{2}\left(\dfrac{1}{a^2} \partial_0\big(a^2\big) \right) \\
&= \frac{3}{2}\left(\dfrac{2\dot a a}{a^2}\right) \\
&= 3\dfrac{\dot a}{a} \\
\\
\Gamma^\alpha_{\; \mu 0} &=
\frac{1}{2}g^{\alpha \kappa}
\left( \partial_\mu g_{\kappa 0} + 
\partial_0 g_{\mu \kappa} -
\partial_\kappa g_{\mu 0}\right) \\
&= \frac{1}{2} g^{\alpha \kappa} (\partial_0 g_{\mu \kappa}) \\
&= \frac{1}{2}\left(\delta^{\alpha \kappa}a^{-2}\right)
\left(\delta_{\mu \kappa} 2\dot a a\right) \\
&= \delta^\alpha_{\; \mu} \dfrac{\dot a}{a}
\end{align*}\\
So we end up with

\begin{align*}
 0 &= 
 -\dot \rho -
 \Gamma^\alpha_{\mu 0} T^{\mu}_{\; \alpha} +
 \Gamma^\mu_{\; \mu 0} T^0_{\; 0} \\
 &= -\dot \rho -
 \delta^\alpha_{\; \mu} \left(\dfrac{\dot a}{a}\right)T^{\mu}_{\; \alpha} +
 3\left(\dfrac{\dot a}{a}\right)  T^0_{\; 0} \\
  &= -\dot \rho - 
  H T^\mu_{\; \mu} +
  3H T^0_{\; 0}\\
  &= \dot \rho +3H(\rho + p)
\end{align*}\\
since

\begin{align*}
&\ T^0_{\; 0} = -\rho \\
&\ T^\mu_{\; \mu} = 3p \quad\text{for $\mu \neq 0$}
\end{align*}
This should be convincing enough that equation \eqref{continuity} represents indeed a continuity equation.

\section{Evolution of FRW universe}

The Friedmann equations \eqref{Friedmann3} and \eqref{Friedmann4}
compose a system of 2 independent equations and are the only independent equations that derive from the Einstein field equations. The acceleration equation \eqref{acceleration equation} and the continuity equation \eqref{continuity} both arise from the Friedmann equations, thus they are not independent of them. However, for a given value of the curvature K (0,1,-1   resulting to different geometries for the universe), the unknown variables that appear in the Friedmann equations are 3; the scale factor (and derivatives of it), the density and pressure. So in order to analytically solve them we would need to introduce a separate, independent equation. This comes in the form of an \emph{equation of state}, a connection between the density and the pressure under the assumption that the universe to behaves like a barotropic fluid, namely that the pressure is a function of only density and vice versa. The simplest one is a linear expression
\begin{equation}\label{equation of state}
p=w\rho
\end{equation}
where $w$ is a constant.\\

\hspace{-0.7cm}So we can now get the picture of the evolution of the universe for each case of spatial geometry for the universe (flat,open or closed). Firstly, we can rewrite the equation \eqref{Friedmann1} as

\begin{equation}
\left(\dfrac{\dot a}{a}\right)^2 -
\dfrac{8\pi G}{3}\rho =
 - \dfrac{K}{a^2}
\end{equation}\\
In that manner we can read this equation as an energy equation, justified by the fact that it derives from the $ "00" $ components of the energy-momentum tensor. The scale factor term
can be associated with the energy that causes the expansion of the energy, while the density term with that of gravitational pull. Furthermore, thanks to the  $"\text{--}" $, we can see that there is a critical density where the 2 term cancel each other out

\begin{equation}\label{critical density}
\dfrac{8\pi G}{3}\rho_{crit} =
\left(\dfrac{\dot a}{a}\right)^2 
\quad\text{or}\quad
\rho_{crit} =
\dfrac{3H^2}{8 \pi G}
\end{equation}  \\
in the same spirit we consider the density parameter as

\begin{equation}\label{Omega}
\Omega \equiv \frac{\rho}{\rho_{crit}}
\end{equation}\\
and by using that notation the 1st Friedmann equation \eqref{Friedmann1}        takes the form

\begin{equation}
\dfrac{K}{H^2 a^2} = \Omega - 1
\end{equation}\\
this form makes it clear to see how the curvature $K$ is directly connected with the components of the universe. We know that $K$ can only take the values 0,1,-1  so, given that the denominator is inherently positive, we are left with only 3 cases

\begin{align*}
&\ \Omega=1 \; (\rho =\rho_{crit}) \Rightarrow
K = 0 \; \Rightarrow  \; \text{flat universe} \\
\\
&\ \Omega>1 \; (\rho > \rho_{crit}) \;\Rightarrow
K = 1 \; \Rightarrow  \; \text{closed universe} \\
\\
&\ \Omega<1 \; (\rho < \rho_{crit}) \;\Rightarrow
K = -1 \; \Rightarrow  \; \text{open universe} \\
\end{align*}\\
We notice that a flat geometry should always remain flat. Additionally, observational results \cite{Hinshaw:2012aka,deBernardis:2000sbo} have shown us that the current value of $\Omega$ is really close to 1 so the spatial geometry of the universe should also be really close to a flat geometry. We will later connect this with one of the problems of the standard cosmological model; namely the \emph{flatness problem}.\\
\\
\hspace{-0.7cm}Working in the case of K=0 we can rewrite the Robertson-Walker metric as so

\begin{equation}
g_{\mu\nu}=diag(-1,a^2,a^2,a^2)
\end{equation}\\

\hspace{-0.7cm}and the Friedmann equations are just

\begin{align*}
&\ H^2 =
\dfrac{8\pi G}{3}\rho  \\
\\ 
&\ \dot H = - 4\pi G (\rho + p) = - 4\pi G \rho (1 + w)
\end{align*}\\
by combining these we get

\begin{align*}
\dfrac{\dot H}{H^2} =
-\frac{3}{2}(1+w)
\Rightarrow
\dfrac{dH}{H^2} =
-\frac{3}{2}(1+w)dt
\end{align*}\\
from that we can calculate the expression for the Hubble parameter to be

\begin{equation}
H=\dfrac{2}{3(1+w)(t-t_0)}\text{,$ \quad w \neq -1$}
\end{equation}\\
where $t_0$ is a constant. Given that the Hubble parameter can be expressed as $H=\dfrac{\dot a}{a}$, we can figure the evolution of the scale factor to be

\begin{equation}
a(t)\varpropto t^{\frac{2}{3(1+w)}}
\end{equation}\\
then the continuity equation \eqref{continuity} gives us the evolution for the density

\begin{equation}
\rho \varpropto a^{-3(1+w)}
\end{equation}\\
both the scale factor and the density are dependent on the constant $w$ which \\ represents the composure of the components on the universe.\\
\\
\hspace{-0.7cm} The most typical examples are the ones were we assume the universe to be dominated by either non-relativistic matter (dust) or radiation, meaning relativistic particles.
In the case of dust we assume an equation of state $w=0$, since non-relativistic matter should have zero pressure. In that case the scale factor and density evolve as

\begin{align*}
&\ a(t) \varpropto t^{\frac{2}{3}}\\
\\
&\ \rho \varpropto a^{-3}
\end{align*}

\hspace{-0.7cm}A universe that evolves in that manner is called \emph{matter-dominated}. In the case of radiation the equation state can be proven to be $w=\frac{1}{3}$ \;(see (8.25) and (8.26) of \cite{Carroll:1997ar}\;). So in that case we get
\begin{align*}
&\ a(t) \varpropto t^{\frac{1}{2}}\\
\\
&\ \rho \varpropto a^{-4}
\end{align*}\\
this corresponds to a \emph{radiation-dominated} universe. These results are not counterintuitive. In the case of ordinary matter we expect the density to be inversely proportional to the volume, since the mass remains always constant, and for an expanding universe the volume is proportional to $a^3$. The same can be applied for radiation, but if we consider relativistic particles to behave as waves, then their wavelength should also be affected by the expansion of the universe by a factor equal to the scale factor $a$. Then since their energy is inversely proportional to the wavelength we expect the density to behave as $\rho \varpropto a^{-4}$.\\

\hspace{-0.7cm}If the universe started off with matter being dominant then the scale factor would evolve as $a(t) \varpropto t^{\frac{2}{3}}$ and the density of matter would evolve as $\rho_m \varpropto a^{-3} \varpropto t^{-2}$ 
where the density of radiation as 
$\rho_r \varpropto a^{-4} \varpropto t^{-\frac{8}{3}}$ and so matter would remain dominant.  If we consider a radiation dominance in the early universe, a more possible scenario considering that the early universe is characterized by high temperatures, the scale factor evolves as $a(t) \varpropto t^{\frac{2}{3}}$ and the corresponding densities as $\rho_m \varpropto a^{-\frac{3}{2}} \varpropto t^{-2}$ , 
$\rho_r \varpropto a^{-4} \varpropto t^{-2}$. So matter should become dominant at some point in the evolution of the universe, a result that is also not counterintuitive as we observe way more non-relativistic matter than radiation, in the form of galaxies and dust clouds.\\

\hspace{-0.7cm}
Both of those cases result in a singularity for $t=0$ named \emph{the Big-Bang} \cite{Mitton:2020wkb}. However both of those solution pose a problem which has to do with the expansion rate of the universe. From the acceleration equation \eqref{acceleration equation} we can see that $\ddot a > 0$ only when $\rho + 3p < 0$. So the acceleration occurs for $w < -\frac{1}{3}$ , which corresponds to a negative pressure. The latest observational evidence dictate that our universe is expanding in an accelerating way \cite{Peebles:2002gy,Filippenko:2000vq,Riess:1998cb,Schmidt:1998ys,Perlmutter:1998np,Bahcall:1999xn}, so we are left with the hypothesis of a $ "vacuum-energy"$ that came to be known as \emph{dark energy}. This comes out of an equation of state $w=-1$ since in that case the energy momentum tensor is 

\begin{align}\label{vaucum energy-momentum}
T_{\mu\nu(\Lambda)} = \lambda \; g_{\mu\nu}
\end{align}\\
where $\lambda$ is a constant. This results to the continuity equation to be just

\begin{align*}
\dot \rho = 0 \;\; \Rightarrow \;  \rho = const.
\end{align*}\\
and by applying this result to the Friedmann equation we get

\begin{align*}
H^2=-\dfrac{8 \pi G}{3}\rho \;\; \Rightarrow \; H=const.
\end{align*}\\
also since the Hubble parameter can be expressed in regard to the scale factor we can once again calculate the evolution of the scale factor

\begin{align*}
H = \dfrac{\dot a}{a} \Rightarrow 
\dfrac{da}{a} = Hdt \Rightarrow
a(t) \varpropto e^{Ht}
\end{align*}\\
a universe dominated by dark energy is called \emph{vacuum-dominated} or \emph{dark energy-dominated} and this result of exponential expansion is called \emph \emph{de-Sitter universe}. The same result can be obtained if we assume a cosmological constant to the Einstein equations.
\\
\\
\\
\\
\\

\section{The Cosmological constant model}

We assume the Einstein field equations with a cosmological constant \eqref{Lambda}. Then the Friedmann equations then yield to

\begin{align}
&\ H^2 - \dfrac{8\pi G}{3}\rho - \dfrac{\Lambda}{3} =
- \dfrac{K}{a^2} \\ \nn
\\
&\ \dfrac{\ddot a}{a} =
-\dfrac{4\pi G}{3}(\rho +3p)+\dfrac{\Lambda}{3}
\end{align}\\
As mentioned Einstein introduced the cosmological constant in an attempt to construct a model of static universe out of his theory \cite{Dadhich:2004pz}. This would mean that $\dot a=0 \; \; and \; \; \ddot a =0$. In the case of matter-dominated universe, meaning $p=0$, we get 

\begin{align*}
\rho = \dfrac{\Lambda}{4\pi G} \quad\text{and}\quad
\Lambda = \dfrac{K}{a^2}
\end{align*}\\
for a matter-dominated universe $\rho > 0$ so $\Lambda$ has to be positive as well, meaning $K=1$, leading to a closed universe. Einstein had to abandon this idea in the 1930's though, as Hubble's observations dictated an expanding universe. Furthermore this static solution can mathematically be proven to be unstable \cite{Eddington:1930zz}. Oddly enough the same mathematics can be used in order to explain the accelerating expansion of the universe, by connecting the cosmological constant with the dark energy. In a flat universe dominated by dark energy the equation of state is $w=-1$ since for vacuum we havethe energy momentum tensor to be proportional to the metric \eqref{vaucum energy-momentum}. Then if we consider the Einstein field equations to be:
\begin{equation}
G_{\mu\nu} = 
8\pi G \left( T_{\mu\nu} + T_{\mu\nu(\Lambda)} \right)
\end{equation}
and the energy-momentum tensor of the cosmological constant to be proportional to the metric we get that

\begin{align*}
\lambda = \dfrac{\Lambda}{8\pi G}
\end{align*}\\
and given that $\lambda$ is associated with the energy density we get 

\begin{align}\label{vacuum density}
\rho_{\scriptscriptstyle \Lambda}=\dfrac{\Lambda}{8\pi G}
\end{align}\\
the energy density appears to be constant, meaning that it is not affected by the expansion of the universe, and its pressure turns out to be negative. Furthermore for a vacuum-dominated the Friedmann equation results in

\begin{align*}
H^2 = 
\dfrac{8\pi G}{3}\rho_{\scriptscriptstyle \Lambda} =
\dfrac{\Lambda}{3} \ra 
a(t) \varpropto e^{\sqrt{\frac{\Lambda}{3}}\; t}
\end{align*}\\
ending up with a similar exponential evolution for the expansion of the universe as we expected for a vacuum-dominated universe. Additionally we can write down the density parameter for the cosmological constant as

\begin{equation}
\Omega_{\scriptscriptstyle \Lambda} \equiv
\dfrac{\rho_{\scriptscriptstyle \Lambda}}{\rho_{crit}} =
\dfrac{\Lambda}{3H^2}
\end{equation}\\
and so the first Friedmann equation can be written generalized as

\begin{equation}\label{Lambda Friedmann}
\ H^2 = \dfrac{8\pi G}{3}\left(\rho +\rho{\scriptscriptstyle \Lambda}\right) -
 \dfrac{K}{a^2} \quad \ra \quad
\Omega + \Omega_{\scriptscriptstyle \Lambda} - 1 =
\dfrac{K}{a^2 H^2}
\end{equation}\\

\section{The standard model of cosmology}

We can now establish the standard cosmological model of Big Bang or \lcdm . CDM comes from Cold Dark Matter (cold refers to low velocities) and $\Lambda$ implies the existence of a cosmological constant. The standard model assumes that the theory of General Relativity is correct and that the universe is composed of ordinary matter, dark matter (introduced in order to explain the rotation of galaxies) and dark energy in the form of a cosmological constant (introduced in order to explain the accelerating expansion). The corresponding parameters of the standard model as observed by the Plank 2018 \cite{Aghanim:2018eyx} are:

\begin{align*}
&\ \bullet \text{baryonic density parameter: \qquad \qquad
 $\Omega_{b_0}$ = 0.0486$\pm$0.0010 } \\
\\
&\ \bullet \text{cold dark matter density parameter: \quad
 $\Omega_{c_0}$ = 0.2589$\pm$0.0057 } \\
\\
&\ \bullet \text{overall matter density parameter: \qquad
$\Omega_{m_0}$ = 0.3089$\pm$0.0062 } \\
\\
&\ \bullet \text{dark energy density parameter: \qquad \;\;
 $\Omega_{\Lambda}$ = 0.6911$\pm$0.0062 } \\
\\
&\ \bullet \text{critical density[kg $m^{-3}$]: \qquad \; \quad\qquad
 $\rho_{0}$ = (8.62$\pm$0.12)$\times 10^{−27}$ } \\
\\
&\ \bullet \text{Hubble constant [ km $s^{-1}$ $Mpc^{-1}$]: \;\;\;\;
 $H_0$ = 67.74$\pm$0.46 } \\
\end{align*}
also the radiation density parameter is estimated to be significantly smaller,\\ according to CMB observations
\begin{align*}
\Omega_{r_0} \sim 5 \times 10^{-5}
\end{align*}
Furthermore we have assumed that the value of the scale factor in the present day is 1. We can make this assumption since the Friedmann equations allow us a freedom to adjust the scale factor  $a \rightarrow \kappa a$ , where $\kappa$ is a constant.
So we can see that the very early universe was dominated by radiation, the energy of the particles is also extremely high. As the universe expands and begins to cool the universe starts to become more matter-dominated, the CMB is the proof of the independence of radiation from matter (decoupling). Today the universe has come to be dominated by dark energy, explaining the accelerated expansion. \\

\hspace{-0.7cm}The standard model of the Big Bang is successful at explaining our universe as we observe it and has also made some significant predictions. Some of those are:
\begin{align*}
\hspace{-3.5cm}&\ \text{$\bullet$ The expansion of our universe} \\
\\
\hspace{-3.5cm}&\ \text{$\bullet$ The existance and spectrum of the Cosmic Microwave Background radiation (CMB)} \\
\\
\hspace{-3.5cm}&\ \text{$\bullet$ The age of the observable universe} 
\\
\\
\hspace{-3.5cm}&\ \text{$\bullet$ The ratios of the lightest atoms in the universe such as \;  H , D , ${}^3 He$ , ${}^4 He$ and ${}^7 Li$} \\
&\ \;\; \text{ as well as the 3 different types (or flavors) of neutrinos.} 
\\
\end{align*}
However, there are several things that cannot be explained within the standard model. Some of the most important problems that have arose are: \\
\\
$\bullet$ \underline{the flatness problem}: \\
 \hspace*{1.1cm}The standard model does not predict a perfectly flat universe, instead \\ \hspace*{1cm}
 the curvature $K$ is a number really close to zero, but not zero. We can \\ \hspace*{1cm} 
 see  from equation \eqref{Lambda Friedmann} that the case of $K=0$ requires  $ \Omega_m + \Omega_{\Lambda} = 1$. \\ \hspace*{1cm} But considering that the early universe was radiation-dominated, the  \\ \hspace*{1cm}
  denominator evolves as $ a^2 H^2 \varpropto t^{-1}$ meaning that $\mid \Omega_m + \Omega_{\Lambda} - 1\mid \varpropto t$ . \\ \hspace*{1cm}
 So the case of $K=0$ is unstable. However given the age of our universe \\ \hspace*{1cm} 
 any small deviation from the critical density that we observe today would \\ \hspace*{1cm}
mean an extremely smaller one at the beginning of the universe. Given that \\ \hspace*{1cm}
 the universe could have any initial conditions for its components, the result \\ \hspace*{1cm}
 that the density should be almost identical to the critical density is strange.\\
 \\
$\bullet$ \underline{the horizon problem}:  \\ \hspace*{1cm}
Observations of the CMB show us inhomogeneities for the overall \\ \hspace*{1cm} 
temperature of the universe to be of a factor of $10^{-5}$, meaning that the \\ \hspace*{1cm}
 universe appears to be in thermodynamic equilibrium. However, since the \\ \hspace*{1cm}
speed that information can travel can not be greater than the speed of light,\\ \hspace*{1cm}
there is  a finite distance that information can travel in order to causally \\ \hspace*{1cm}
connect two points in the universe. This distance is the \emph{horizon}. We can \\ \hspace*{1cm}
calculate that the horizon for the CMB corresponds to a distance of $1^o$ in \\ \hspace*{1cm}
the sky. So the CMB should not appear to be isotropic.
\\
\\
$\bullet$ \underline{the monopole problem}:  \\ \hspace*{1cm}
Grand unified theories that describe the unification of the fundamental  \\ \hspace*{1cm}
 forces predict the existence of magnetic monopoles and other supersym-  \\ \hspace*{1cm}
 metric particles. These are created at the very early stages of the universe  \\ \hspace*{1cm}
 and should in theory dominate the present universe. However as of today  \\ \hspace*{1cm}
 we have been unable to observe any of those particles.
 \\
 \\
 \\
 \\\\\\
 $\bullet$ \underline{the fine tuning problem of the cosmological constant}:  \\ \hspace*{1cm}
Observational evidence shows that the vacuum energy density is compera-\\ \hspace*{1cm}
ble  to the critical density, approximately $\rho_{\scriptscriptstyle \Lambda} \simeq 0.7 \rho_0 \simeq 10^{-47} \;GeV^4$.  \\ \hspace*{1cm}
However quantum field theory estimates the vacuum energy to be
\\ \hspace*{1cm} 
$\rho_{\scriptscriptstyle \Lambda} \simeq 10^{74} \; GeV^4$
leading to a difference of 121 orders of magnitude. So if
\\ \hspace*{1cm} the cosmological constant is indeed connected to the vacuum energy then \\ \hspace*{1cm}
it has to be very precisely adjusted to a really small, non-zero value, similar \\ \hspace*{1cm}
fine tuning problem as with the flatness of the universe.
 \\\\
In order to address these problems we have to look for some extensions for the standard model. The most noteworthy is the one that was proposed at the 1980's by Alan Guth as well as Alexei Starobinsky and  Andrei Linde called \emph{cosmological inflation}.

\chapter{Dark energy models}
\section{Inflation}

The basic idea of inflation is that at a very early stage the universe begun to rapidly expand \cite{Liddle:1999mq,Guth:1980zm,Guth:1982ec,Senatore:2016aui,Sriramkumar:2009kg,Linde:1981mu,Linde:1983gd,DeFelice:2010aj,Lola:2020lvk}. This rapid expansion is connected to a vacuum energy-dominated era and one of the simpler models to describe it is the cosmological constant, which leads to an exponential increase for the scale factor .

 This immediately solves the flatness problem, since any initial value for the curvature parameter K that is not significantly big will lead to a current value of zero. It is believed that the inflation begun when the universe was of age $10^{-36}$ sec
 and ended at $10^{-32}$ sec. In this small amount the universe 
could have been increased in size up to roughly $10^{47}$ times.

 Furthermore it also solves the horizon problem since the short period of time that is required is sufficient to maintain thermodynamic equilibrium and at the end of the inflation the entire universe would be causally connected. And it can remain like that until the decoupling where the CMB originates.
 
 Finally inflation solves the monopole problem as well, since after the inflation the density of magnetic monopoles would drastically decrease and the universe would become radiation-dominated, meaning that they would be extremely hard to be observed today.
 
 However inflation itself does not solve the fine tuning problem of the cosmological constant. For that we have to assume that the value of the cosmological constant is indeed zero and that dark energy can be explained via other mechanisms. In this section we will examine some other dark energy models\cite{Copeland:2006wr,Chevallier:2000qy,Bamba:2012cp,Ferreira:1997hj,Sahni:2002fz,Lymperis:2018iuz}.

\section{Quintessence}

Quintessence \cite{Carroll:1998zi,Ratra:1987rm,Caldwell:1997ii,Lymperis:2017ulc,Lymperis:2018ajw} is an attempt of describing dark energy with the help of a scalard field $\phi$, as a scalar field can indeed lead to negative pressure. The corresponding action should be

\begin{equation}\label{Qaction}
S_{\phi} =
\int \sqrt{-g}
\left(-\frac{1}{2}g^{\alpha \beta}\partial_\alpha \phi \partial_\beta \phi - 
V(\phi)\right)d^4x
\end{equation}\\
The term $\; g^{\alpha \beta}\partial_\alpha \phi \partial_\beta = (\nabla \phi)^2 \;$ refers to the kinetic term and $V(\phi )$ is the potential of the scalar field. The variation of this action leads us to an equation of motion similarly to the derivation of the Einstein field equations.

\begin{align*}
\delta  S_{\phi} =
 - \int &\ \bigg\{\; \delta \big(\sqrt{-g}\big) \left(\frac{1}{2}g^{\alpha \beta}\partial_\alpha \phi \partial_\beta \phi +V(\phi) \right)\\
  &+ \sqrt{-g} \; \delta \left(\frac{1}{2}g^{\alpha \beta}\partial_\alpha \phi \partial_\beta \phi +V(\phi) \right) \; \bigg\} \; d^4x \\
  = - \int &\ \bigg\{\; \delta \big(\sqrt{-g}\big) \left(\frac{1}{2}g^{\alpha \beta}\partial_\alpha \phi \partial_\beta \phi +V(\phi) \right)\\
  &+ \frac{1}{2}\sqrt{-g} \;\delta \left(g^{\alpha \beta}\partial_\alpha \phi \partial_\beta \phi \right) + \sqrt{-g}\; \delta \big(V(\phi) \big)\; \bigg\} \; d^4x
  \\
\end{align*}
by calculating the variation of the expressions that come up we get 

\begin{align*}
&\ \delta \big(\sqrt{-g}\big) =
- \frac{1}{2} \sqrt{-g} \; g_{\mu\nu} \delta g^{\mu\nu} \\
\\
&\ \delta \big(V(\phi) \big) =
 V(\phi +\delta \phi) - V(\phi) =
\dfrac{\partial V}{\partial \phi}\delta \phi + O(\delta {\phi}^2)
 \simeq \dfrac{\partial V}{\partial \phi} \delta \phi \\
 \\
&\ \delta \left(g^{\alpha \beta }\partial_\alpha \phi \partial_\beta \phi \right) =
\partial_\alpha \phi \partial_\beta \phi \delta  g^{\alpha \beta }+
g^{\alpha \beta } \delta \big(\partial_\alpha \phi \partial_\beta \phi \big)
\end{align*}\\
and for the term $\delta \big(\partial_\alpha \phi \partial_\beta \phi \big)$ we use a similar product rule

\begin{align*}
g^{\alpha \beta } \delta \big(\partial_\alpha \phi \partial_\beta \phi \big) &=
g^{\alpha \beta }\delta(\partial_\alpha \phi)\partial_\beta \phi  +
g^{\alpha \beta }\delta(\partial_\beta \phi)\partial_\alpha \phi  \\
&= g^{\alpha \beta }\partial_\alpha(\delta  \phi)\partial_\beta \phi +
g^{\alpha \beta }\partial_\beta(\delta  \phi)\partial_\alpha \phi \\
&= 2g^{\alpha \beta } \partial_\alpha \phi \partial_\beta \delta \phi
\\
\end{align*}
Furthermore we can notice that 

\begin{align*}
\sqrt{-g}\; g^{\alpha \beta } \partial_\alpha \phi \;\partial_\beta \delta \phi = 
\partial_\beta\;\big(\sqrt{-g}\; g^{\alpha \beta } \partial_\alpha \phi \; \delta \phi \big) -
\partial_\beta\;\big(\sqrt{-g}\; g^{\alpha \beta } \partial_\alpha \phi \big) \; \delta \phi \\
\end{align*}
where $\partial_\beta\;\big(\sqrt{-g}\; g^{\alpha \beta } \partial_\alpha \phi \; \delta \phi \big)$ leads to boundary terms after the integration, so as we have seen before we can ignore those terms considering boundaries that go to infinity. Thus we get the following from for the variation of the action
\\
\\
\\

\begin{align}\label{scalar field action}
\delta  S_{\phi} =
 -\int &\ \bigg\{- \frac{1}{2} \sqrt{-g} \; g_{\mu\nu} \left(\frac{1}{2}g^{\alpha \beta}\partial_\alpha \phi \partial_\beta \phi +V(\phi) \right) \delta g^{\mu\nu}\nn \\ 
 &+ \; \frac{1}{2} \sqrt{-g}\; \bigg(\partial_\alpha \phi \partial_\beta \phi\bigg) \delta  g^{\alpha \beta } \\
&+ \bigg(-\partial_\beta\;\big(\sqrt{-g}\; g^{\alpha \beta } \partial_\alpha \phi \big)  +\sqrt{-g}\; \dfrac{\partial V}{\partial \phi} \bigg) 
\delta \phi\;\bigg\} d^4x \nn \\ \nn
\end{align}
and from the principle of least action for the variation over the scalar field leads to

\begin{align*}
\dfrac{\delta S_\phi}{\delta \phi} = 0 \ra 
- \partial_\beta \big(\sqrt{-g}\; g^{\alpha \beta} \partial_\alpha \phi \big) +
\sqrt{-g}\dfrac{\partial V}{\partial \phi} =0
\end{align*} \\
for a flat FRW universe the metric is \eqref{FRW flat} and its determinant is $g=-a^6$. So the derivative becomes

\begin{align*}
\hspace{-2cm} \partial_\beta \big(\sqrt{-g}\; g^{\alpha \beta} \partial_\alpha \phi \big) &=  \;\;
 g^{\alpha \beta}\partial_\alpha \phi \; \partial_\beta \sqrt{-g} \\
&\ \; + \sqrt{-g}\;\partial_\beta g^{\alpha \beta} \;\partial_\alpha \phi \\
&\ \; + \sqrt{-g}\; g^{\alpha \beta}\partial_\beta \partial_\alpha \phi
\end{align*}\\
and by calculating the expressions that came up we get

\begin{align*}
g^{\alpha \beta}\partial_\alpha \phi \; \partial_\beta \big(\sqrt{-g} \big) &= 
g^{00}\partial_0 \phi \; \partial_0 \sqrt{-g}  + 
g^{ii}\partial_i \phi \; \partial_i \sqrt{-g}  
\qquad\text{$\big(\;\partial_i \sqrt{-g} =0\big)$} \\
&= -3a^2\dot a \;\dot \phi \\
\\
\sqrt{-g}\;\partial_\beta g^{\alpha \beta} \;\partial_\alpha \phi &=
0 \qquad\text{$\big(\partial_\beta g^{\alpha \beta} =0 \big)$}\\
\\
\sqrt{-g}\; g^{\alpha \beta}\partial_\beta \partial_\alpha \phi &=
a^3\big( -\ddot \phi + \dfrac{1}{a^2}{\nabla}^2 \phi \big)
\end{align*} \\\
Hence the equation of motion for the scalar field takes the form

\begin{equation}\label{K-G equation}
\ddot \phi + 3H\dot \phi -\dfrac{1}{a^2}{\nabla}^2 \phi + 
\dfrac{\partial V}{\partial \phi} =0
\end{equation}\\
the term $3H\dot \phi \;$ can be seen as a friction term for the momentum of the field due to the expansion. Furthermore we can consider the field to be smooth across space, without any significant fluctuations. So comparing to the time derivative we can ignore the spatial derivatives, then the above equation is just

\begin{equation}\label{K-G equation 2}
\ddot \phi + 3H\dot \phi + 
\dfrac{\partial V}{\partial \phi} =0
\end{equation}\\
If we were to add the action \eqref{scalar field action} to the action we used in order to derive the Einstein field equations for the case of matter
\begin{align*}
S_{tot}= 
S_{\scriptscriptstyle H-E} +
S_{\scriptscriptstyle Matter} +
S_\phi
\end{align*}\\
the variation over the inverse metric would lead us to a similar result

\begin{equation}
R_{\mu\nu} -\frac{1}{2}g_{\mu\nu}R = 
8\pi G \bigg(T_{\mu\nu}^{(Matter)} + T_{\mu\nu}^{(\phi)}\bigg)
\end{equation}\\
where we get an extra source term in the form of an energy-momentum part that refers to the scalar field. The definition for the energy-momentum tensor would also be similar

\begin{equation}
T_{\mu\nu}^{(\phi)} \equiv
- \dfrac{2}{\sqrt{-g}}
\dfrac{\delta S_{\phi}}{\delta g^{\mu\nu}}
\end{equation}\\
and we end up with the following expression 

\begin{equation}
T_{\mu\nu}^{(\phi)} = 
\partial_\mu \phi \partial_\nu \phi -
 g_{\mu\nu} \left(\frac{1}{2}g^{\alpha \beta}\partial_\alpha \phi \partial_\beta \phi +V(\phi) \right)
\end{equation}\\
from that we can calculate the corresponding energy density and pressure for the scalar field

\begin{align*}
\rho_\phi = 
 T^0_{\;0} = g^{00} T_{00} =
\frac{1}{2} \dot {\phi}^2 -
 \frac{1}{2a^2}\big(\nabla \phi \big)^2 +
 V(\phi)\\
 \\
 p_\phi = \frac{1}{3} T^i_{\;i} = \frac{1}{3}g^{ij} T_{ij} =
 \frac{1}{2} \dot {\phi}^2 -
  \frac{1}{6a^2}\big(\nabla \phi \big)^2 -
 V(\phi)\\
\end{align*}
where we can again ignore the spatial derivatives and get a simpler form for the energy density and pressure

\begin{align}
\rho_\phi = 
\frac{1}{2} \dot {\phi}^2  +
 V(\phi)\\ \nn
 \\
 p_\phi = 
 \frac{1}{2} \dot {\phi}^2 -
 V(\phi)\\ \nn
\end{align}\\
So the generalized Friedmann equations for the case of a flat universe are just 

\begin{align}
&\ H^2 =
\dfrac{8\pi G}{3}\;\big(\rho + \rho_{\phi}\big)\label{GFriedmann1}\\ \nonumber
\\ 
&\ \dfrac{\ddot a}{a}= -\dfrac{4\pi G}{3}\;\big[(\rho + \rho_{\phi}) +3(p+p_{\phi})\big]\label{GFriedmann2}
\end{align}\\
and for a scalar field-dominated universe those equations yield to

\begin{align}
&\ H^2 =
\dfrac{8\pi G}{3}\;\bigg(\frac{1}{2}\dot \phi^2 + V(\phi)\bigg)\label{GFriedmann3} \\ \nn
\\
&\ \dfrac{\ddot a}{a} =- \dfrac{8\pi G}{3}\;\bigg(\dot \phi^2 - V(\phi)\bigg)\label{GFriedmann4}
\end{align}\\
we can consider and equation of state

\begin{equation}
w_{\phi}=
\dfrac{p_{\phi}}{\rho_{\phi}}=
\dfrac{\dot \phi^2 - 2V(\phi)}{\dot \phi^2 + 2V(\phi)}
\end{equation}\\
where $w_{\phi}$ is not generally constant. Given that the continuity equation becomes

\begin{equation}
\dot \rho_{\phi} +3H(\rho_{\phi} + p_{\phi})= 0
\end{equation}\\
and has a similar solution for the energy density, the only difference being that  $w_{\phi}$ can be , generally, dependent of the scale factor

\begin{equation}
 \rho_{\phi} = 
 \rho_{\phi}(t_0)\;e^{-\int 3(1+w_{\phi})\frac{da}{a}}
 \end{equation}\\
in the quintessence model we assume the potential energy to be far greater than the kinetic energy of the field. In that limit the field satisfies the condition

\begin{equation}\label{potential condition}
 \dot \phi^2 << V(\phi)
 \end{equation} \\
so the equation of state falls into the vacuum-dominated FRW universe category of $w_{\phi} = -1 $ with

\begin{equation}
\rho_{\phi} \simeq -V(\phi) \quad\text{and}\quad 
p_{\phi} \simeq V(\phi)
\end{equation}\\
the continuity equation then dictates that the energy pressure of the scalar field, and consequently the pressure, would be constant, meaning that in that limit the potential tends to become constant as well. Furthermore in that limit the Friedmann equation yields to

\begin{equation}
H_{\phi}^2 \simeq \dfrac{8 \pi G}{3} V(\phi)
\end{equation}\\
thus a constant Hubble parameter, as we already know, leads to the following exponential growth for the scale factor (de Sitter phase)

\begin{equation}
a(t) \varpropto e^{H_{\phi}(t-t_0)} 
\end{equation}\\
The hypothesis that the potential energy is more potent than the kinetic energy does not come without any basis. That is to assume a "flat" scalar field where we ignore the term $\ddot \phi$ with the idea that even if this term started off with a high value, the pressure term $3H\dot \phi $ from the Klein-Gordon equation \eqref{K-G equation 2} would sooner or later make its value significantly smaller. In that case equation \eqref{K-G equation 2} becomes

\begin{equation}
 \dot \phi = -\dfrac{1}{3H}\partial_{\phi}V
 \end{equation}\\
in order to achieve an inflationary scenario the potential must be close to constant. So the term $\partial_{\phi}V$ has to be really small and thus we get the condition \eqref{potential condition}.


\section{K-essence}

In Quintessence model we had the requirement of the condition \eqref{potential condition}, meaning that the inflation occurred due to the potential energy of a scalar field. However there are inflationary models that instead, focus more so on the kinetic energy named \emph{K-inflation} \cite{ArmendarizPicon:1999rj,ArmendarizPicon:2000dh}, the most general form of which includes non-canonical terms. The corresponding dark energy models are called \emph{K-essence} \cite{Chiba:1999ka,ArmendarizPicon:2000ah,}.

The Lagrangian that describes those models, in its general form, is

\begin{equation}\label{Kaction}
S_{\kappa} = \int \sqrt{-g}\;p(\phi,X)\;d^4x
\end{equation}\\
where the Lagrangian density $p$ reflects purely on the pressure terms and is a function of the field $\phi$ and the kinetic term
 $X=-\frac{1}{2}(\nabla \phi)^2=-\frac{1}{2}g^{\alpha\beta}\;\partial_\alpha \phi \;\partial_\beta \phi$.
 
We can derive the same action as in Quintessence \eqref{Qaction} if we elect an energy pressure for the action \eqref{Kaction} of the form

\begin{equation}\label{Qpressure}
p(\phi,X)= -\frac{1}{2}(\nabla \phi)^2 - V(\phi)
\end{equation}\\
so in that sense Quintessence is just a sub-case of K-essence. However
K-essence models that have the most applications are those that focus purely on the kinetic terms and ignore any potential-related contribution. These require the energy pressure to be of the form

\begin{equation}
p(\phi,X)=f(\phi)p(X) \text{,}\qquad\text{$f(\phi)>0$}
\end{equation}\\
which falls under the category of \emph{purely kinetic K-essence}. That requirement excludes any potential-oriented scenarios and consequently the Quintessence model as it can be easily seen that the energy pressure \eqref{Qpressure} used in Quintessence does not satisfy the above condition.

Considering only the contribution of kinetic terms and including non-canonical ones we get the action

\begin{equation}
S_X = \int \sqrt{-g} \;\big( A(\phi)X + B(\phi)X^2 + ... \big)\;d^4x
\end{equation}\\
and for small values of $X$ we can ignore quadratic and higher order terms, so by keeping only the terms up to the second order the Lagrangian density is just 

\begin{equation}
p(\phi,X) =  A(\phi)X + B(\phi)X^2
\end{equation}\\
which falls under the category of K-essence if we consider a transformation for the field

\begin{equation}
\phi_{old} \rightarrow \phi_{new}= \int_{\phi_{old}} d\phi \sqrt{\dfrac{B}{\mid A \mid}}
\end{equation}\\
then the Lagrangian density takes the form 

\begin{equation}
p(\phi,X) = f(\phi)\big(-X + X^2\big)
\end{equation}\\
the corresponding energy-momentum tensor is

\begin{equation}
T_{\mu\nu}=
2\dfrac{\partial p(\phi,X)}{\partial X}\partial_\mu \phi \; \partial_\nu \phi - p(\phi,X) g_{\mu\nu}
\end{equation}\\
and so for the energy density we get

\begin{equation}
\rho_\phi =
2X \dfrac{\partial p}{\partial X} - p =
f(\phi)\big(-X+X^2\big)
\end{equation}\\
the equation of state then is a function of only $X$

\begin{equation}\label{Kstate}
w_\phi=
\dfrac{p_\phi}{\rho_\phi}=
\dfrac{1-X}{1-3X}
\end{equation}\\
furthermore for the case of a flat FRW universe the Friedmann equations are

\begin{align}
&H^2 =
\dfrac{8 \pi G}{3}f(\phi)\big(-X+X^2\big) \\ \nn
\\
&\dfrac{\ddot a}{a}= -\dfrac{4 \pi G}{3}\rho_\phi(1+3w_{\phi})
\end{align}\\
so the accelerated expansion occurs for $w_\phi < -\frac{1}{3}$ which corresponds to values of $X< \frac{2}{3}$. Additionally the case of a cosmological constant, meaning $w_\phi = -1$, corresponds to the value $X=\frac{1}{2}$.

\section{Phantom field}

The case with an equation of state  $w<-1$ is usually described by a scalar field that is characterized by negative kinetic energy and is known as \emph{phantom field} \cite{Carroll:2003st,Caldwell:1999ew,Singh:2003vx,Sami:2003xv}. This is also supported by observational data \cite{Corasaniti:2004sz} and can not only give us an inflationary scenario, but can also explain the current acceleration of the universe, since an energy form with equation of state $w<-1$ affects the universe in its much later states, given the evolution equations of a flat FRW universe.  

In order to achieve the negative kinetic energy we take the opposite sign on the kinetic term in the action \eqref{Qaction}. So we get

\begin{equation}
S_{ph} =
\int \sqrt{-g}
\left(\frac{1}{2}g^{\alpha \beta}\partial_\alpha \phi \partial_\beta \phi - 
V(\phi)\right)d^4x
\end{equation}\\
This leads to the following energy-density and energy-pressure values

\begin{align}\label{Ph1}
\rho_{ph} = 
-\frac{1}{2} \dot {\phi}^2  +
 V(\phi)\\ \nn
 \\
 p_{ph} = 
 -\frac{1}{2} \dot {\phi}^2 -
 V(\phi)\\ \nn
\end{align}\\
and to the equation of state 

\begin{equation}\label{PhEqofstate}
w_{ph}=
\dfrac{p_{ph}}{\rho_{ph}}=
\dfrac{\dot \phi^2 + 2V(\phi)}{\dot \phi^2 - 2V(\phi)}
\end{equation}\\
which allows the case of $w<-1$ for $\;\dot \phi^2 < 2V(\phi)$.\\
\\
\\

\hspace{-0.7cm}The equations of motion for an early phantom field dominated universe are

\begin{align}
&\ H^2 =
\dfrac{8\pi G}{3}\;\bigg(-\frac{1}{2}\dot \phi^2 + V(\phi)\bigg)\label{GFriedmann3p} \\ \nn
\\
&\ \dfrac{\ddot a}{a} =- \dfrac{8\pi G}{3}\;\bigg(-\dot \phi^2 - V(\phi)\bigg)\label{GFriedmann4p}\\ \nn
\\
&\ddot \phi + 3H\dot \phi + 
\dfrac{\partial V}{\partial \phi} =0
\end{align}\\

Similarly to Quintessence this model can predict the accelerated rate of expansion. However this case of $w<-1$ leads inevitably to what is called \emph{the Big Rip} \cite{Caldwell:2003vq}, a singularity where the scale factor diverges after a finite amount of time. For that, a more generalized form of the phantom field can be introduced, one inspired by the K-essence models, with non-canonical kinetic terms being introduced. This allows for the existence of a theory that predicts an energy form with negative kinetic energy and given the equation of state \eqref{Kstate} can see that these scenarios are acceptable.

\chapter{Epilogue}

In this thesis we have examined only but a few possible scenarios that can justify and describe dark energy, but there is an even larger family of models that tries to accomplish the same. The wide range of those models can be due to our lack of understanding the very nature of dark energy and thus the effort for such a theory to emerge is continuous, with some believing that the answer lies with \emph{Quantum Gravity} \cite{Sorkin:1997gi} or \emph{Supergravity}\cite{Bailin:1994qt,Nilles:1983ge,Weinberg:1988cp,Ellis:1982wr}. However this road seems to be long and hard which makes it one of the most exciting problems in our generation. 

\section{Aknowledgements}

I want to thank my supervisor Prof. Smaragda Lola for introducing me to this topic and my Co-advisor PhD Student Andreas Lymperis for helping me throughout the construction of this thesis, as well as the Andreas Mentzelopoulos foundation for their financial support.

\bibliographystyle{plain}
\bibliography{bibliography}

\end{document}